\documentclass[review]{elsarticle}
\journal{Signal Processing}

\usepackage{amsfonts,amsmath,amssymb}
\usepackage{pifont}
\usepackage{enumitem}
\usepackage{array}
\usepackage{multirow}
\usepackage{graphicx}
\usepackage{subfig}
\graphicspath{{./images/}}
\usepackage{lscape}
\usepackage{algorithm}
\usepackage{algpseudocode}
\usepackage{enumitem}

\newtheorem{definition}{Definition}
\newtheorem{remark}{Remark}
\newtheorem{cost}{Cost function}
\newtheorem{pena}{Penalty}

\newcommand{\norm}[1]{\left\lVert#1\right\rVert}
\newcounter{mymodel}
\newenvironment{mymodel}{\refstepcounter{mymodel}\equation}{\tag{M\themymodel}\endequation}
\newcounter{myproblem}
\newenvironment{myproblem}{\refstepcounter{myproblem}\equation}{\tag{P\themyproblem}\endequation}
\newcounter{mycost}
\newenvironment{mycost}{\refstepcounter{mycost}\equation}{\tag{C\themycost}\endequation}
\newcommand\one{\mathds{1}}
\usepackage{dsfont}
\newcommand{\sca}[2]{\left \langle #1 \middle| #2 \right \rangle}
\newcommand{\pa}[1]{\lfloor #1 \rfloor}
\newcommand{\mycite}[1]{\citeauthor{#1} (\citeyear{#1})}

 \usepackage{setspace}

\biboptions{numbers,sort&compress}

\begin{document}

\begin{frontmatter}
\title{Selective review of offline change point detection methods}

\author[mymainaddress]{Charles Truong\corref{mycorrespondingauthor}}
\author[mysecondaryaddress]{Laurent Oudre}
\author[mymainaddress]{Nicolas Vayatis}

\address[mymainaddress]{CMLA, CNRS, ENS Paris Saclay}
\address[mysecondaryaddress]{L2TI, University Paris 13}

\cortext[mycorrespondingauthor]{Corresponding author}
\begin{abstract}
	\noindent
	This article presents a selective survey of algorithms for the offline detection of multiple change points in multivariate time series.
	A general yet structuring methodological strategy is adopted to organize this vast body of work.
	More precisely, detection algorithms considered in this review are characterized by three elements: a cost function, a search method and a constraint on the number of changes.
	Each of those elements is described, reviewed and discussed separately.
	Implementations of the main algorithms described in this article are provided within a Python package called ruptures.
\end{abstract}
\begin{keyword}
change point detection, segmentation, statistical signal processing
\end{keyword}

\end{frontmatter}

\section{Introduction}

A common task in signal processing is the identification and analysis of complex systems whose underlying state changes, possibly several times. This setting arises when industrial systems, physical phenomena or human activity are continuously monitored with sensors. The objective of practitioners is to extract from the recorded signals a posteriori meaningful information about the different states and transitions of the monitored object for analysis purposes. This setting encompasses a broad range of real-world scenarios and a wide variety of signals.

Change point detection is the task of finding changes in the underlying model of a signal or time series.
The first works on change point detection go back to the 50s~\cite{Page1954,Page1955}: the goal was to locate a shift in the mean of independent and identically distributed (iid) Gaussian variables for industrial quality control purposes.
Since then, this problem has been actively investigated, and is periodically the subject of in-depth monographs~\cite{Basseville1993,Darkhovsky1993,Csorgo1997,chen2011parametric}.
This subject has generated important activity in statistics and signal processing~\citep{Lavielle2007, Jandhyala2013, Haynes2017} but also in various application settings such as speech processing~\cite{desobry2005online,Harchaoui2009,Angelosante2012,Seichepine2014}, financial analysis~\cite{Lavielle2007,Bai1998a,Frick2014}, bio-informatics~\cite{Hocking2013,Maidstone2017,Vert2010,Picard2005,guedon:inria-00311634,Chakar2017,Oudre2015a,Audiffren2015,Liu2018}, climatology~\cite{Maidstone2013,Verbesselt2010,Reeves2007}, network traffic data analysis~\cite{Levy-Leduc2009,Lung-Yut-Fong2012}.
Modern applications in bioinformatics, finance, monitoring of complex systems have also motivated recent developments from the machine learning community~\citep{Vert2010,Lajugie2014,Hocking2015}.

\begin{figure}[t]
	\centering
	\includegraphics[width=0.8\columnwidth]{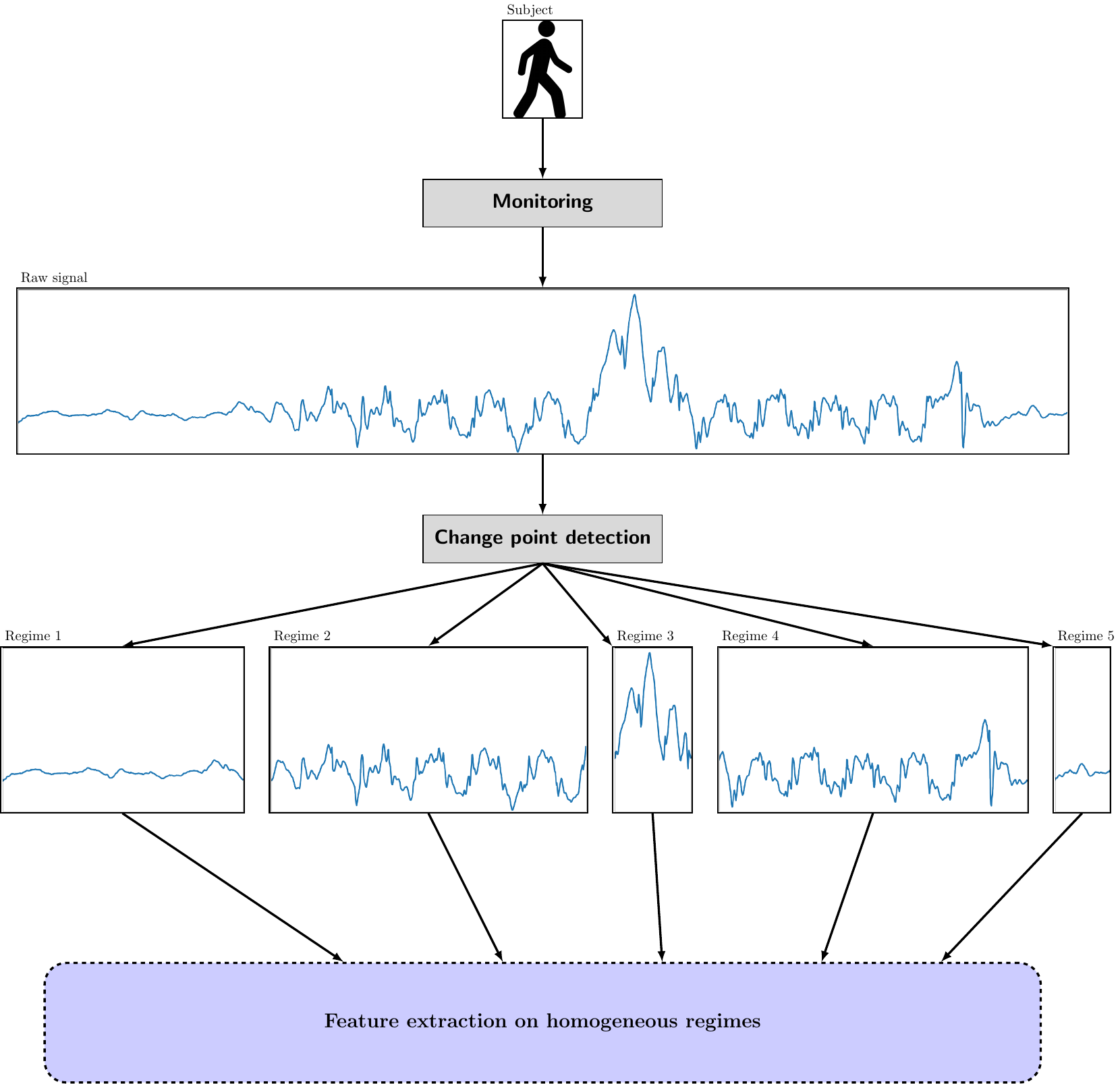}
	\caption{Flowchart of a study scheme, for gait analysis.}
	\label{fig:intro_marche}
\end{figure}%

Let us take the example of gait analysis, illustrated on the flowchart displayed on~Figure~\ref{fig:intro_marche}. In this context, a patient's movements are monitored with accelerometers and gyroscopes while performing simple activities, for instance walking at preferred speed, running or standing still. The objective is to objectively quantify gait characteristics~\cite{Barrois-Muller2016a,Barrois-Muller2016,s18114033,Truong2015,Barrois-Muller2015}. The resulting signal is described as a succession of non-overlapping segments, each one corresponding to an activity and having its own gait characteristics. Insightful features from homogeneous phases can be extracted if the temporal boundaries of those segments are identified. This analysis therefore needs a preliminary processing of the signals: change point detection.

Change point detection methods are divided into two main branches: \textit{online} methods, that aim to detect changes as soon as they occur in a real-time setting, and \textit{offline} methods that retrospectively detect changes when all samples are received. The former task is often referred to as \textit{event or anomaly detection}, while the latter is sometimes called \textit{signal segmentation}.

In this article, we propose a survey of algorithms for the detection of multiple change points in multivariate time series. All reviewed methods presented in this paper address the problem of \textit{offline} (also referred to as \textit{retrospective} or \textit{a posteriori}) change point detection, in which segmentation is performed after the signal has been collected. The objective of this article is to facilitate the search of a suitable detection method for a given application. In particular, focus is made on  practical considerations such as implementations and procedures to calibrate the algorithms. This review also presents the mathematical properties of the main approaches, as well as the metrics to evaluate and compare their results. This article is linked with a Python scientific library called \texttt{ruptures} \cite{packrup}, that includes a modular and easy-to-use implementation of all the main methods presented in this paper.

\section{Background}

This section introduces the  main concepts for change point detection, as well as the selection criteria and the outline of this review.

\subsection{Notations}
In the remainder of this article, we use the following notations.
For a given signal $y=\{y_t\}_{t=1}^T$, the $(b-a)$-sample long sub-signal $\{y_t\}_{t=a+1}^{b}$ ($1\leq a < b \leq T$) is simply denoted $y_{a..b}$; the complete signal is therefore $y=y_{0..T}$.
A set of indexes is denoted by a calligraphic letter: $\mathcal{T} = \{t_1,t_2,\dots\}\subset \{1,\dots,T\} $, and its cardinal is $|\mathcal{T}|$.
For a set of indexes $\mathcal{T}=\{t_1,\dots,t_K\}$, the dummy indexes $t_0:=0$ and $t_{K+1}:=T$ are implicitly available.

\subsection{Problem formulation}

Let us consider a multivariate non-stationary random process $y = \{ y_1 ,\dots , y_T \} $ that takes value in $\mathbb{R}^d$ ($d \geq 1$) and has $T$ samples.
The signal $y$ is assumed to be piecewise stationary, meaning that some characteristics of the process change abruptly at some unknown instants $t^{*}_1 < t^{*}_2  < \dots < t^{*}_{K^{*}}$.
Change point detection consists in estimating the indexes $t^{*}_k$.
Depending on the context, the number $K^{*}$ of changes may or may not be known, in which case it has to be estimated too.

Formally, change point detection is cast as a model selection problem, which consists in choosing the best possible segmentation $\mathcal{T}$ according to a quantitative criterion $V(\mathcal{T},y)$ that must be minimized.
(The function $V(\mathcal{T}, y)$ is simply denoted $V(\mathcal{T})$ when it is obvious from the context that it refers to the signal $y$.)
The choice of the criterion function $V(\cdot)$ depends on preliminary knowledge on the task at hand.

In this work, we make the assumption that the criterion function $V(\mathcal{T})$ for a particular segmentation is a sum of costs of all the segments that define the segmentation:
\begin{equation}
	V(\mathcal{T}, y) := \sum_{k=0}^{K} \ c(y_{t_{k}..t_{k+1}})
	\label{eq:review_sum_of_cost}
\end{equation}
where $c(\cdot)$ is a cost function which measures goodness-of-fit of the sub-signal $y_{t_{k}..t_{k+1}}=\{y_t\}_{t_k+1}^{t_{k+1}}$ to a specific model.
The ``best segmentation'' $\widehat{\mathcal{T}}$ is the minimizer of the criterion $V(\mathcal{T})$.
In practice, depending on whether the number $K^{*}$ of change points is known beforehand, change point detection methods fall into two categories.

\begin{itemize}
\item \textbf{Problem 1\label{problem-kn}~: known number of changes $K$.} The change point detection problem with a fixed number $K$ of change points consists in solving the following discrete optimization problem
	      \begin{myproblem}\label{eq:review_kn}
		      \min_{|\mathcal{T}|=K}\ V(\mathcal{T}).
	      \end{myproblem}
          \item \textbf{Problem 2\label{problem-unkn}~: unknown number of changes.} The change point detection problem with an unknown number of change points consists in solving the following discrete optimization problem
	      \begin{myproblem}\label{eq:review_unkn}
		      \min_{\mathcal{T}}\ V(\mathcal{T}) \ +\ \mbox{pen} (\mathcal{T})
	      \end{myproblem}
	      where $\mbox{pen}(\mathcal{T})$ is an appropriate measure of the complexity of a segmentation $\mathcal{T}$.
          
\end{itemize}

All change point detection methods considered in this work yield an exact or an approximate solution to either Problem~1~\eqref{eq:review_kn} or Problem~2~\eqref{eq:review_unkn}, with the function $V(\mathcal{T}, y)$ adhering to the format~\eqref{eq:review_sum_of_cost}.

\subsection{Selection criteria for the review}
\begin{figure}[t]
	\centering
	\includegraphics[width=0.6\columnwidth]{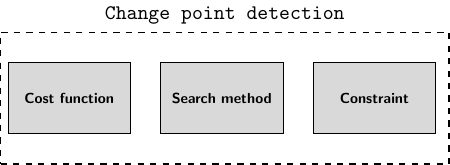}
	\caption{Typology of change point detection methods described in this article.
		Reviewed algorithms are defined by three elements: a cost function, a search method and a constraint (on the number of change points).
	}
	\label{fig:review-diagram}
\end{figure}
\noindent
To better understand the strengths and weaknesses of change point detection methods, we propose to classify algorithms according to a comprehensive typology.
Precisely, detection methods are expressed as the combination of the following three elements.
\begin{itemize}
	\item \textbf{Cost function.}\enspace
	      The cost function $c(\cdot)$ is a measure of ``homogeneity''.
	      Its choice encodes the type of changes that can be detected.
	      Intuitively, $c(y_{a..b})$ is expected to be low if the sub-signal $y_{a..b}$ is ``homogeneous'' (meaning that it does not contain any change point), and large if the sub-signal $y_{a..b}$ is ``heterogeneous'' (meaning that it contains one or several change points).
	\item \textbf{Search method.}\enspace
	      The search method is the resolution procedure for the discrete optimization problems associated with Problem~1~\eqref{eq:review_kn} and Problem~2~\eqref{eq:review_unkn}.
	      The literature contains several methods to efficiently solve those problems, in an exact fashion or in an approximate fashion.
	      Each method strikes a balance between computational complexity and accuracy.
	\item \textbf{Constraint (on the number of change points).}\enspace
	      When the number of changes is unknown~\eqref{eq:review_unkn}, a constraint is added, in the form of a complexity penalty $\mbox{pen}(\cdot)$~\eqref{eq:review_unkn}, to balance out the goodness-of-fit term $V(\mathcal{T},y)$.
	      The choice of the complexity penalty is related to the amplitude of the changes to detect: with too ``small'' a penalty (compared to the goodness-of-fit) in~\eqref{eq:review_unkn}, many change points are detected, even those that are the result of noise.
	      Conversely, too much penalization only detects the most significant changes, or even none.
\end{itemize}
This typology of change point detection methods is schematically shown on Figure~\ref{fig:review-diagram}.

\subsection{Limitations}
The described framework, however general, does not encompass all published change point detection methods.
In particular, Bayesian approaches are not considered in the remainder of this article, even though they provide state-of-the-art results in several domains, such as speech and sound processing.
The most well-known Bayesian algorithm is the Hidden Markov Model (HMM)~\cite{Rabiner1989}.
This model was later extended, for instance with Dirichlet processes~\cite{Ko2015,Martinez2014} or product partition models~\cite{Barry1992, Barry1993}.
The interested reader can find reviews of Bayesian approaches in~\cite{Darkhovsky1993} and~\cite{chen2011parametric}.\\
Also, several literature reviews with different selection criteria can be found.
Recent and important works include~\cite{Aminikhanghahi2017} which focuses on window-based detection algorithms.
In particular, the authors use the quantity of samples needed to detect a change as a basis for comparison.
Maximum likelihood and Bayes-type detection are reviewed, from a theoretical standpoint, in~\cite{Jandhyala2013}.
Existing asymptotic distributions for change point estimates are described for several statistical models.
In~\cite{Niu2016}, detection is formulated as a statistical hypothesis testing problem, and emphasis is put on the algorithmic and theoretical properties of several sequential mean-shift detection procedures.

\subsection{Outline of the article}

Before starting this review, we propose in Section \ref{sec:eval} a detailed overview of the main mathematical tools that can be used for evaluating and comparing the change point detection methods. The organization of the remaining of this review article reflects the typology of change point detection methods, which is schematically shown on Figure~\ref{fig:review-diagram}.
Precisely, the three defining elements of a detection algorithm are reviewed separately.
In Section~\ref{sec:review_costs}, cost functions from the literature are presented, along with the associated signal model and the type of change that can be detected.
Whenever possible, theoretical results on asymptotic consistency are also given.
Section~\ref{sec:review_kn} lists search methods that efficiently solve the discrete optimizations associated with Problem~1~\eqref{eq:review_kn} and Problem~2~\eqref{eq:review_unkn}.
Both exact and approximate methods are described.
Constraints on the number of change points are reviewed in Section~\ref{sec:review_unkn}.
A summary table of the literature review can be found in Section~\ref{sec:review_summary}.
The last section \ref{sec:python} is dedicated to the presentation of the Python package that goes with this article and propose a modular implementation of all the main approaches described in this article.

\section{Evaluation}
\label{sec:eval}

Change point detection methods can be evaluated either by proving some mathematical properties of the algorithms (such as consistency) in general case, or empirically by computing several metrics to assess the performances on a given dataset.

\subsection{Consistency}
\label{sec:review-asymptotic}
A natural question when designing detection algorithms is the consistency of estimated change point indexes, as the number of samples $T$ goes to infinity.
In the literature, the ``asymptotic setting'' is intuitively described as follows: the observed signal $y$ is regarded as a realization of a continuous-time process on an equispaced grid of size $1/T$, and ``$T$ goes to infinity'' means that the spacing of the sampling grid converges to 0.
Precisely, for all $\tau\in[0,1]$, let $Y(\tau)$ denote an $\mathbb{R}^d$-valued random variable such that
\begin{equation}
	y_t = Y(t/T) \quad \forall t=1,\dots, T.
\end{equation}
The continuous-time process undergoes $K^{*}$ changes in the probability distribution at the time instants $\tau^{*}_k \in(0,1)$.
Those $\tau^{*}_k$ are related to the change point indexes $t^{*}_k$ through the following relationship:
\begin{equation}
	t^{*}_k = \lfloor T\tau^{*}_k\rfloor.
\end{equation}
Generally, for a given change point index $t_k$, the associated quantity $\tau_k = t_k/T \in (0,1)$ is referred to as a change point \emph{fraction}.
In particular, the change point fractions $\tau^{*}_{k}$ ($k=1,\dots,K^{*}$) of the time-continuous process $Y$ are change point indexes of the discrete-time signal $y$.
Note that in this asymptotic setting, the lengths of each regime of $y$ increase linearly with $T$.
The notion of asymptotic consistency of a change point detection method is formally introduced as follows.
\begin{definition}[Asymptotic consistency]\label{def:review_consistance}
	A change point detection algorithm is said to be asymptotically consistent if the estimated segmentation $\widehat{\mathcal{T}} = \{\hat{t}_1,\hat{t}_2,\dots\}$ satisfies the following conditions, when $T\longrightarrow +\infty$:
	\begin{enumerate}[label=(\roman*)]
		\item $P(|\widehat{\mathcal{T}}|=K^{*})\longrightarrow1$,
		\item $\frac{1}{T}\norm{\widehat{\mathcal{T}}-\mathcal{T}^*}_{\infty} \overset{p}{\longrightarrow} 0$,
	\end{enumerate}
	where the distance between two change point sets is defined by
	\begin{equation}
		\norm{\widehat{\mathcal{T}}-\mathcal{T}^*}_{\infty} := \max\ \{\ \max_{\hat{t}\in\widehat{\mathcal{T}}}\ \min_{ t^{*} \in\mathcal{T}^*} |\hat{t}- t^{*} |,\ \max_{ t^{*} \in\mathcal{T}^*} \min_{\hat{t}\in\widehat{\mathcal{T}}} |\hat{t}- t^{*} |\ \}.
	\end{equation}
\end{definition}

	In Definition~\ref{def:review_consistance}, the first condition is trivially verified when the number $K^{*}$ of change points is known beforehand.
	As for the second condition, it implies that the estimated change point fractions are consistent, and not the indexes themselves.
	In general, distances $|\hat{t} - t^{*} |$ between true change point indexes and their estimated counterparts do not converge to 0, even for simple models~\cite{Bai2004,Chakar2014,Vert2010,Boysen2009}.
	As a result, consistency results in the literature only deal with change point fractions.

\subsection{Evaluation metrics}
\label{sec:Evaluation}
Several metrics from the literature are presented below.
Each metric correspond to one of the previously listed criteria by which segmentation performances are assessed.
In the following, the set of true change points is denoted by $\mathcal{T}^*= \{t^*_1,\dots,t^*_{K^*} \} $, and the set of estimated change points is denoted by $\widehat{\mathcal{T}}= \{\hat{t}_1,\dots,\hat{t}_{\widehat{K}} \}$.
Note that that the cardinals of each set, $K^*$ and $\widehat{K}$, are not necessarily equal.
\subsubsection{AnnotationError}
The \textsc{AnnotationError} is simply the difference between the predicted number of change points $|\widehat{\mathcal{T}}|$ and the true number of change points $|\mathcal{T}^{\star}|$:
\begin{equation}
	\textsc{AnnotationError} := |\widehat{K}-K^*|.
\end{equation}
This metric can be used to discriminate detection method when the number of changes is unknown.
\subsubsection{Hausdorff}
\begin{figure}[t]
	\centering
	\includegraphics[width=0.6\columnwidth]{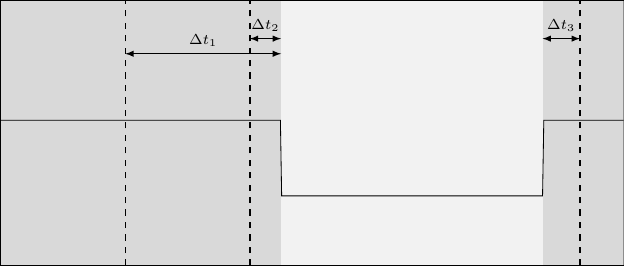}
	\caption{\textsc{Hausdorff}.
		Alternating gray areas mark the segmentation $\mathcal{T}^*$; dashed lines mark the segmentation $\widehat{\mathcal{T}}$.
		Here, \textsc{Hausdorff} is equal to $\Delta t_1 = \max (\Delta t_1,\Delta t_2,\Delta t_3)$.
	}
	\label{fig:hausdorff}
\end{figure}

The \textsc{Hausdorff} metric measures the robustness of detection methods~\cite{Boysen2009, Harchaoui2010}.
Formally, it is equal to the greatest temporal distance between a change point and its prediction:
\begin{equation*}
	\textsc{Hausdorff}( \mathcal{T}^*, \widehat{\mathcal{T}} ) := \max\ \{\ \max_{\hat{t}\in\widehat{\mathcal{T}}}\ \min_{ t^* \in\mathcal{T}^*} |\hat{t}- t^* |,\ \max_{ t^* \in\mathcal{T}^*} \min_{\hat{t}\in\widehat{\mathcal{T}}} |\hat{t}- t^* |\ \}.
	\label{eq:corpus_hausdorff}
\end{equation*}%
It is the worst error made by the algorithm that produced $\widehat{\mathcal{T}}$ and is expressed in number of samples.
If this metric is equal to zero, both breakpoint sets are equal; it is large when a change point from either $\mathcal{T}^*$ or $\widehat{\mathcal{T}}$ is far from every change point of $\widehat{\mathcal{T}}$ or $\mathcal{T}^*$ respectively.
Over-segmentation as well as under-segmentation is penalized.
An illustrative example is displayed on Figure~\ref{fig:hausdorff}.
\subsubsection{RandIndex}
\begin{figure}[t]
	\centering
	\includegraphics[width=0.6\columnwidth]{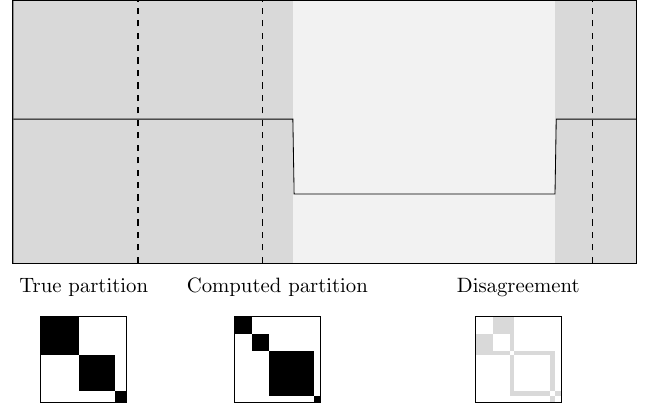}
	\caption{\textsc{RandIndex}.
		Top: alternating gray areas mark the segmentation $\mathcal{T}^*$; dashed lines mark the segmentation $\widehat{\mathcal{T}}$.
		Below: representations of associated adjacency matrices and disagreement matrix.
		The adjacency matrix of a segmentation is the $T\times T$ binary matrix with coefficient $(s,t)$ equal to 1 if $s$ and $t$ belong to the same segment, 0 otherwise.
		The disagreement matrix is the  $T\times T$ binary matrix with coefficient $(s,t)$ equal to 1 where the two adjacency matrices disagree, and 0 otherwise.
		\textsc{RandIndex} is equal to the white area (where coefficients are 0) of the disagreement matrix.
	}
	\label{fig:RandIndex}
\end{figure}

Accuracy can be measured by the \textsc{RandIndex}, which is the average similarity between the predicted breakpoint set $\widehat{\mathcal{T}}$ and the ground truth $\mathcal{T}^*$~\cite{Lajugie2014}.
Intuitively, it is equal to the number of agreements between two segmentations.
An agreement is a pair of indexes which are either in the same segment according to both $\widehat{\mathcal{T}}$ and $\mathcal{T}^*$ or in different segments according to both $\widehat{\mathcal{T}}$ and $\mathcal{T}^*$.
Formally, for a breakpoint set $\mathcal{T}$, the set of grouped indexes and the set of non-grouped indexes are respectively $\mathrm{gr}(\mathcal{T})$ and $\mathrm{ngr}(\mathcal{T})$:
\begin{equation*}
	\begin{split}
		\mathrm{gr}(\mathcal{T}) &:= \{ (s, t), 1\leq s<t\leq T \text{ s.t. } s \text{ and } t \text{ belong to the same segment according to } \mathcal{T}\ \}, \\
		\mathrm{ngr}(\mathcal{T}) &:= \{ (s, t), 1\leq s<t\leq T \text{ s.t. } s \text{ and } t \text{ belong to different segments according to } \mathcal{T}\ \}.
	\end{split}
\end{equation*}%
The \textsc{RandIndex} is then defined as follows:
\begin{equation}
	\text{\textsc{RandIndex}}(\mathcal{T}^*,\widehat{\mathcal{T}}) := \frac{|\mathrm{gr}(\widehat{\mathcal{T}}) \cap\mathrm{gr}(\mathcal{T}^*)  | + | \mathrm{ngr}(\widehat{\mathcal{T}}) \cap \mathrm{ngr}(\mathcal{T}^*)|)}{T(T-1)} .
	\label{eq:corpus_randindex}
\end{equation}%
It is normalized between 0 (total disagreement) and 1 (total agreement).
Originally, \textsc{RandIndex} has been introduced to evaluate clustering methods~\cite{Boysen2009,Lajugie2014}.
An illustrative example is displayed on Figure~\ref{fig:RandIndex}.
\subsubsection{F1-score}
\begin{figure}[t]
	\centering
	\includegraphics[width=0.6\columnwidth]{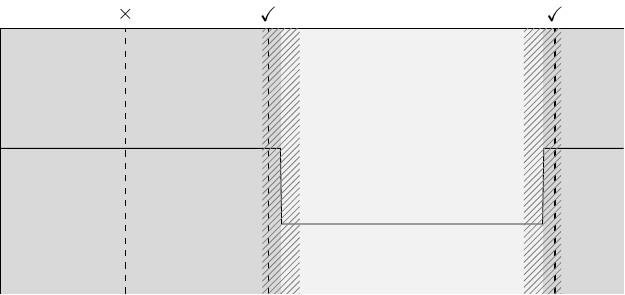}
	\caption{\textsc{F1-Score}.
		Alternating gray areas mark the segmentation $\mathcal{T}^*$; dashed lines mark the segmentation $\widehat{\mathcal{T}}$; dashed areas mark the allowed margin of error around true change points.
		Here, \textsc{Prec} is $2/3$, \textsc{Rec} is $2/2$ and \textsc{F1-Score} is $4/5$.
	}
	\label{fig:precision}
\end{figure}

Another measure of accuracy is the \textsc{F1-Score}.
Precision is the proportion of predicted change points that are true change points.
Recall is the proportion of true change points that are well predicted.
A breakpoint is considered detected up to a user-defined margin of error $M>0$; true positives {\sc Tp} are true change points for which there is an estimated one at less than $M$ samples, \emph{i.e.}
\begin{equation}
	\text{\sc Tp}(\mathcal{T}^*,\widehat{\mathcal{T}}):= \{t^*\in\mathcal{T}^*\ |\ \exists\, \hat{t}\in\widehat{\mathcal{T}}\ \ \text{s.t.}\ |\hat{t} - t^*|<M \}.
\end{equation}%
Precision \textsc{Prec} and recall \textsc{Rec} are then given by
\begin{equation}
	\text{{\sc Prec}}(\mathcal{T}^*,\widehat{\mathcal{T}}):=|\text{\sc Tp}(\mathcal{T}^*,\widehat{\mathcal{T}})|/\widehat{K} \quad \text{and}\quad
	\text{{\sc Rec}}(\mathcal{T}^*,\widehat{\mathcal{T}}):=|\text{\sc Tp}(\mathcal{T}^*,\widehat{\mathcal{T}})|/K^*.
	\label{eq:corpus_precrec}
\end{equation}
\textsc{Precision} and \textsc{Recall} are well-defined (ie. between 0 and 1) if the margin $M$ is smaller than the minimum spacing between two true change point indexes $t^*_k$ and $t^*_{k+1}$.
Over-segmentation of a signal causes the precision to be close to zero and the recall close to one.
Under-segmentation has the opposite effect.
The \textsc{F1-Score} is the harmonic mean of precision {\sc Prec} and recall {\sc Rec}:
\begin{equation}
	\text{\textsc{F1-Score}}(\mathcal{T}^*,\widehat{\mathcal{T}}) := 2\times\frac{\textsc{Prec}(\mathcal{T}^*,\widehat{\mathcal{T}})\times\textsc{Rec}(\mathcal{T}^*,\widehat{\mathcal{T}})}{\textsc{Prec}(\mathcal{T}^*,\widehat{\mathcal{T}}) + \textsc{Rec}(\mathcal{T}^*,\widehat{\mathcal{T}})}.
	\label{eq:corpus_f1}
\end{equation}
Its best value is 1 and its worse value is 0.
An illustrative example is displayed on Figure~\ref{fig:precision}.

\section{Models and cost functions}\label{sec:review_costs}
\begin{figure}[t]
	\centering
	\includegraphics[width=\columnwidth]{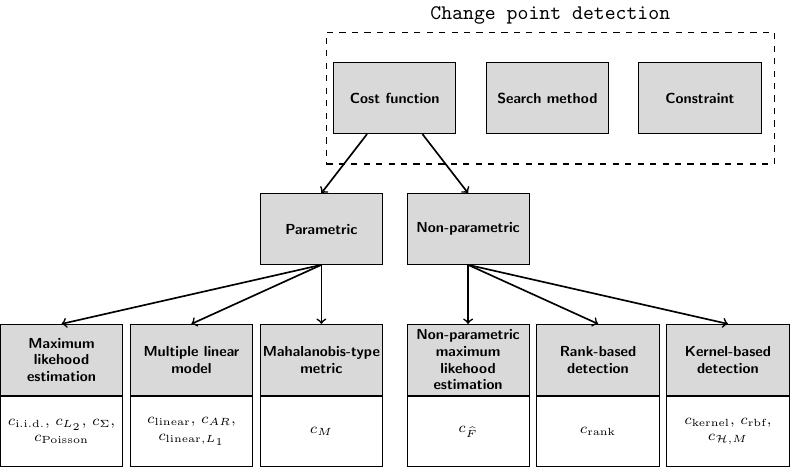}
	\caption{Typology of the cost functions described in Section~\ref{sec:review_costs}.}
	\label{fig:review-diagram-cost}
\end{figure}

This section presents the first defining element of change detection methods, namely the cost function.
In most cases, cost functions are derived from a signal model.
In the following, models and their associated cost function are organized in two categories: parametric and non-parametric, as schematically shown in Figure~\ref{fig:review-diagram-cost}.
For each model, the most general formulation is first given, then special cases, if any, are described.
A summary table of all reviewed costs can be found at the end of this section.

\subsection{Parametric models}
Parametric detection methods focus on changes in a finite-dimensional parameter vector.
Historically, they were the first to be introduced, and remain extensively studied in the literature.

\subsubsection{Maximum likelihood estimation}\label{sec:cost_iid}
Maximum likelihood procedures are ubiquitous in the change point detection literature.
They generalize a large number of models and cost functions, such as mean-shifts and scale shifts in normally distributed data~\cite{Page1955,Lavielle1999,Pein2017,Keshavarz2018}, changes in the rate parameter of Poisson distributed data~\cite{Ko2015}, etc.
In the general setting of maximum likelihood estimation for change detection, the observed signal $y=\{y_1,\dots,y_T\}$ is composed of independent random variables, such that
\begin{mymodel}\label{eq:cost_pw_iid_model}
	y_t \sim \sum_{k=0}^{K^{*}} f(\cdot| \theta_k) \one( t^{*}_k < t \leq t^{*}_{k+1} )
\end{mymodel}
where the $t^{*}_k$ are change point indexes, the $f(\cdot|\theta)$ are probability density functions parametrized by the vector-valued parameter $\theta$, and the $\theta_k$ are parameter values.
In other words, the signal $y$ is modelled by iid variables with piecewise constant distribution.
The parameter $\theta$ represents a quantity of interest whose value changes abruptly at the unknown instants $t^{*}_k$, which are to be estimated.
Under this setting, change point detection is equivalent to maximum likelihood estimation if the sum of cost $V(\mathcal{T},y)$ is equal to the negative log-likelihood.
The corresponding cost function, denoted $c_{\mbox{i.i.d.}}$, is defined as follows.

\begin{cost}[$c_{\mbox{i.i.d.}}$]
	\label{def:review_c_iid}
	For a given parametric family of distribution densities $\{f(\cdot|\theta) | \theta \in\Theta\} $ where $\Theta$ is a compact subset of $\mathbb{R}^p$ (for a certain $p$), the cost function $c_{\mbox{i.i.d.}}$ is defined by
	\begin{mycost}\label{eq:cost_general_likelihood}
		c_{\mbox{i.i.d.}}(y_{a..b}) := - \sup\limits_{\theta} \sum_{t=a+1}^{b} \log f(y_t|\theta).
	\end{mycost}
\end{cost}
\noindent
Model~\ref{eq:cost_pw_iid_model} and the related cost function $c_{\mbox{i.i.d.}}$ encompasses a large number of change point methods.
Note that, in this context, the family of distributions must be known before performing the detection, usually thanks to prior knowledge on the data.
Historically, the Gaussian distribution was first used, to model mean-shifts~\cite{Lavielle2000,Srivastava75binseg,Krishnaiah1988} and scale shifts~\cite{Ko2015,Aue2012,Pein2017}.
A large part of the literature then evolved towards other parametric distributions, most notably resorting to distributions from the general exponential family~\cite{Maidstone2013,Fearnhead2006,Frick2014}.\\
From a theoretical point of view, asymptotic consistency, as described in Definition~\ref{def:review_consistance}, has been demonstrated, in the case of \emph{a single change point}, first with Gaussian distribution (fixed variance), then for several specific distributions, e.g.\ Gaussian with mean and scale shifts~\cite{Basseville1993,chen2011parametric,Keshavarz2018,Gorecki2018}, discrete distributions~\cite{Lavielle1999}, etc.
The case with \emph{multiple change points} has been tackled later.
For certain distributions (e.g.\ Gaussian), the solutions of the change point detection problems~\eqref{eq:review_kn} (known number of change points) and~\eqref{eq:review_unkn} (unknown number of change points) have been proven to be asymptotically consistent~\cite{Fu1990}.
The general case of multiple change points and a generic distribution family has been addressed decades after the change detection problem has been introduced: the solution of the change point detection problem with a known number of changes and a cost function set to $c_{\mbox{i.i.d.}}$ is asymptotically consistent~\cite{He2010}.
This is true if certain assumptions are satisfied: (i) the signal follows the model~\eqref{eq:cost_pw_iid_model} for a distribution family that verifies some regularity assumptions (which are no different from the assumptions needed for generic maximum likelihood estimation, without any change point) and (ii) technical assumptions on the value of the cost function on homogeneous and heterogeneous sub-signals.
As an example, distributions from the exponential family satisfy those assumptions.
\paragraph{Related cost functions.}
The general model~\eqref{eq:cost_pw_iid_model} has been applied with different families of distributions.
We list below three notable examples and the associated cost functions: change in mean, change in mean and scale, and change in the rate parameter of count data.\\

\begin{itemize}
	\item The mean-shift model is the earliest and one of the most studied model in the change point detection literature~\cite{Page1955,Chernoff1964,Lorden1971,Mallows1973,Srivastava75binseg}.
	      Here, the distribution is Gaussian, with fixed variance.
	      In other words, the signal $y$ is simply a sequence of independent normal random variables with piecewise constant mean and same variance.
	      In this context, the cost function $c_{\mbox{i.i.d.}}$ becomes $c_{L_2}$, defined below.
	      This cost function is also referred to as the quadratic error loss and has been applied for instance on DNA array data~\cite{Frick2014} and geology signals~\cite{Chen2011a}.
	      \begin{cost}[$c_{L_2}$]
		      \label{ex:cost_mean_shift}
		      The cost function $c_{L_2}$ is given by
		      \begin{mycost}\label{eq:cost_mean_shift}
			      c_{L_2}(y_{a..b}) := \sum_{t=a+1}^{b} \norm{y_t - \bar{y}_{a..b}}^2_2
		      \end{mycost}
		      where $\bar{y}_{a..b}$ is the empirical mean of the sub-signal $y_{a..b}$.
	      \end{cost}
	\item A natural extension to the mean-shift model consists in letting the variance abruptly change as well.
	      In this context, the cost function $c_{\mbox{i.i.d.}}$ becomes $c_{\Sigma}$, defined below.
	      This cost function can be used to detect changes in the first two moments of random (not necessarily Gaussian) variables, even though it is the Gaussian likelihood that is plugged in $c_{\mbox{i.i.d.}}$~\cite{Lavielle1999,Jandhyala2013}.
	      It has been applied for instance on stock market time series~\cite{Lavielle1999}, biomedical data~\cite{Chen2011a}, and electric power consumption monitoring~\cite{Rossi2010neurocomputing}.
	      \begin{cost}[$c_{\Sigma}$]
		      The cost function $c_{\Sigma}$ is given by
		      \begin{mycost}
			      c_{\Sigma}(y_{a..b}) := (b-a) \log\det\widehat{\Sigma}_{a..b} + \sum_{t=a+1}^b (y_t - \bar{y}_{a..b})' \widehat{\Sigma}_{a..b}^{-1} (y_t - \bar{y}_{a..b})
			      \label{eq:cost_scale_shift}
		      \end{mycost}
		      where $\bar{y}_{a..b}$ and $\widehat{\Sigma}_{a..b}$ are respectively the empirical mean and the empirical covariance matrix of the sub-signal $y_{a..b}$.
	      \end{cost}
	\item Change point detection has also be applied on count data modelled by a Poisson distribution~\cite{Chib1998,Ko2015}.
	      More precisely, the signal $y$ is a sequence of independent Poisson distributed random variables with piecewise constant rate parameter.
	      In this context, the cost function $c_{\mbox{i.i.d.}}$ becomes $c_{\mbox{Poisson}}$, defined below.
	      \begin{cost}[$c_{\mbox{Poisson}}$]
		      The cost function $c_{\mbox{Poisson}}$ is given by
		      \begin{mycost}\label{eq:cost_poisson}
			      c_{\mbox{Poisson}}(y_{a..b}) := - (b-a) \bar{y}_{a..b}\log\bar{y}_{a..b}
		      \end{mycost}
		      where $\bar{y}_{a..b}$ is the empirical mean of the sub-signal $y_{a..b}$.
	      \end{cost}
\end{itemize}

\begin{remark}
	A model slightly more general than~\eqref{eq:cost_pw_iid_model} can be formulated by letting the signal samples to be dependant and the distribution function $f(\cdot|\theta)$ to change over time.
	This can in particular model the presence of unwanted changes in the statistical properties of the signal (for instance in the statistical structure of the noise~\cite{Lavielle1999}).
	The function $f(\cdot|\theta)$ is replaced in~\eqref{eq:cost_pw_iid_model} by a sequence of distribution functions $f_t(\cdot|\theta)$ which are not assumed to be identical for all indexes $t$.
	Changes in the functions $f_t$ are considered nuisance parameters and only the variations of the parameter $\theta$ must be detected.
	Properties on the asymptotic consistency of change point estimates can be obtained in this context.
	We refer the reader to~\cite{Lavielle1999, Lavielle2005} for theoretical results.
\end{remark}%

\subsubsection{Piecewise linear regression}
Piecewise linear models are often found, most notably in the econometrics literature, to detect so-called ``structural changes''~\cite{Bai1994, Bai1995, Bai1996}.
In this context, a linear relationship between a response variable and covariates exists, and this relationship changes abruptly at some unknown instants.
Formally, the observed signal $y$ follows a piecewise linear model with change points located at the $t^{*}_k$:
\begin{mymodel}
	\forall\ t,\ t^{*}_k<t\leq t^{*}_{k+1},\quad y_t = x_t'u_k + z_t'v + \varepsilon_t \quad (k=0,\dots, K^{*})
	\label{eq:cost_linear_model}
\end{mymodel}
where the $u_k\in\mathbb{R}^p$ and $v \in \mathbb{R}^q$ are unknown regression parameters and $\varepsilon_t$ is noise.
Under this setting, the observed signal $y$ is regarded as a univariate response variable (ie $d=1$) and the signals $x=\{x_t\}_{t=1}^{T}$ and $z=\{z_t\}_{t=1}^{T}$ are observed covariates, respectively $\mathbb{R}^p$-valued and $\mathbb{R}^q$-valued.
In this context, change point detection can be carried out by fitting a linear regression on each segment of the signal.
To that end, the sum of costs is made equal to the sum of squared residuals.
The corresponding cost function, denoted $c_{\mbox{linear}}$, is defined as follows.
\begin{cost}[$c_{\mbox{linear}}$]
	For a signal $y$ (response variable) and covariates $x$ and $z$, the cost function $c_{\mbox{linear}}$ is defined by
	\begin{mycost}
		c_{\mbox{linear}}(y_{a..b}) := \min_{u\in\mathbb{R}^p, v\in\mathbb{R}^q}\ \sum_{t=a+1}^{b} (y_t - x_t'u - z_t'v)^2.
		\label{eq:cost_linear}
	\end{mycost}
\end{cost}
\noindent
In the literature, Model~\eqref{eq:cost_linear_model} is also known as a \emph{partial} structural change model because the linear relationship between $y$ and $x$ changes abruptly, while the linear relationship between $y$ and $z$ remains constant.
The \emph{pure} structural change model is obtained by removing the term $z_t'v$ from~\eqref{eq:cost_linear_model}.
This formulation generalizes several well-known models such as the autoregressive (AR) model~\cite{Angelosante2012,Bai2000}, multiple regressions~\cite{Bai1996,Bai1997a}, etc.
A more general formulation of~\eqref{eq:cost_linear_model} that can accommodate a multivariate response variable $y$ exists~\cite{Qu2007}, but is more involved, from a notational standpoint.\\
From a theoretical point of view, piecewise linear models are extensively studied in the context of change point detection by a series of important contributions~\cite{Bai1994, Bai1995, Bai1996, Bai1997a, Bai1998a, Bai1998b, Bai1999, Bai2000, Bai2003, Perron2006, Bai2010}.
When the number of changes is known, the most general consistency result can be found in~\cite{Bai1998a}.
A multivariate extension of this result has been demonstrated in~\cite{Qu2007}.
As for the more difficult situation of an unknown number of changes, statistical tests have been proposed for a single change~\cite{Bai1998} and multiple changes~\cite{Bai1999}.
All of those results are obtained under various sets of general assumptions on the distributions of the covariates and the noise.
The most general of those sets can be found in~\cite{Perron2006a}.
Roughly, in addition to some technical assumptions, it imposes the processes $x$ and $z$ to be weakly stationary within each regime, and precludes the noise process to have a unit root.
\paragraph{Related cost functions.}
In the rich literature related to piecewise linear models, the cost function $c_{\mbox{linear}}$ has been applied and extended in several different settings.
Two related cost functions are listed below.
\begin{itemize}
	\item The first one is $c_{\mbox{linear}, L_1}$, which was introduced in order to accommodate certain noise distributions with heavy tails~\cite{Bai1995,Bai1998b} and is defined as follows.
	      \begin{cost}[$c_{\mbox{linear}, L_1}$]
		      For a signal $y$ (response variable) and covariates $x$ and $z$, the cost function $c_{\mbox{linear}, L_1}$ is defined by
		      \begin{mycost}\label{eq:linear_cost_abs}
			      c_{\mbox{linear}, L_1}(y_{a..b}) := \min_{u\in\mathbb{R}^p, v\in\mathbb{R}^q}\ \sum_{t=a+1}^{b} |y_t - x_t'u - z_t'v|.
		      \end{mycost}
	      \end{cost}
	      \noindent%
	      The difference between $c_{\mbox{linear}, L_1}$ and $c_{\mbox{linear}}$ lies in the norm used to measure errors: $c_{\mbox{linear}, L_1}$ is based on a least absolute deviations criterion, while $c_{\mbox{linear}}$ is based on a least squares criterion.
	      As a result, $c_{\mbox{linear}, L_1}$ is often applied on data with noise distributions with  heavy tails~\cite{Maidstone2013,Jandhyala2013}.
	      In practice, the cost function $c_{\mbox{linear}, L_1}$ is computationally less efficient than the cost function $c_{\mbox{linear}}$, because the associated minimization problem~\eqref{eq:linear_cost_abs} has no analytical solution.
	      Nevertheless, the cost function $c_{\mbox{linear}, L_1}$ is often applied on economic and financial data~\cite{Bai1994,Bai1995,Bai1996}.
	      For instance, changes in several economic parameters of the G-7 growth have been investigated using a piecewise linear model and $c_{\mbox{linear}, L_1}$~\cite{Doyle2005}.
	\item The second cost function related to $c_{\mbox{linear}}$ has been introduced to deal with piecewise autoregressive signals.
	      The autoregressive model is a popular representation of random processes, where each variable depends linearly on the previous variables.
	      The associated cost function, denoted $c_{AR}$, is defined as follows.
	      \begin{cost}[$c_{AR}$]\label{ex:cost_ar}
		      For a signal $y$ and an order $p\geq 1$, the cost function $c_{AR}$ is defined by
		      \begin{mycost}\label{eq:linear_cost_ar}
			      c_{AR}(y_{a..b}) := \min_{u\in\mathbb{R}^p}\ \sum_{t=a+1}^{b} \norm{y_t - x_t'u}^2
		      \end{mycost}
		      where $x_t:=[y_{t-1},y_{t-2},\dots,y_{t-p}]$ is the vector of lagged samples.
	      \end{cost}
	      \noindent
	      The piecewise autoregressive model is a special case of the generic piecewise linear model, where the term $z_t'v$ is removed (yielding a pure structural change model) and the covariate signal $x$ is equal to the signal of lagged samples.
	      The resulting cost function $c_{AR}$ is able to detect shifts in the autoregressive coefficients of a non-stationary process~\cite{Bai2000,Chakar2017}.
	      This model has been applied on EEG/ECG time series~\cite{Qu2007}, functional magnetic resonance imaging (fMRI) time series~\cite{Nam2012} and speech recognition tasks~\cite{Angelosante2012}.
\end{itemize}
\subsubsection{Mahalanobis-type metric}
The cost function $c_{L_2}$~\eqref{eq:cost_mean_shift}, adapted for mean-shift detection, can be extended through the use of Mahalanobis-type pseudo-norm.
Formally, for any symmetric positive semi-definite matrix $M\in\mathbb{R}^{d\times d}$, the associated pseudo-norm $\norm{\cdot}_M$ is given by:
\begin{equation}
	\norm{y_t}^2_M := y_t' M y_t
\end{equation}
for any sample $y_t$.
The resulting cost function $c_M$ is defined as follows.
\begin{cost}[$c_{M}$]
	The cost function $c_{M}$, parametrized by a symmetric positive semi-definite matrix $M\in\mathbb{R}^{d\times d}$, is given by
	\begin{mycost}
		c_M(y_{a..b}) := \sum_{t=a+1}^{b} \norm{y_t - \bar{y}_{a..b}}^2_M
		\label{eq:cost_mahalanobis}
	\end{mycost}
	where $\bar{y}_{a..b}$ is the empirical mean of the sub-signal $y_{a..b}$.
\end{cost}
\noindent
Intuitively, measuring distances with the pseudo-norm $\norm{\cdot}_M$ is equivalent to applying a linear transformation on the data and using the regular (Euclidean) norm $\norm{\cdot}$.
Indeed, decomposing the matrix $M = U'U$ yields:
\begin{equation}
	\norm{y_t - y_s}^2_M = \norm{Uy_t - Uy_s}^2.
\end{equation}
Originally, the metric matrix $M$ was set equal to the inverse of the covariance matrix, yielding the Mahalanobis metric~\cite{Mahalanobis1936}, ie
\begin{equation}
	M = \widehat{\Sigma}^{-1}
\end{equation}
where $\widehat{\Sigma}$ is the empirical covariance matrix of the signal $y$.
By using $c_{M}$, shifts in the mean of the transformed signal can be detected.
In practice, the transformation $U$ (or equivalently, the matrix $M$) is chosen to highlight relevant changes.
This cost function generalizes all linear transformations of the data samples.
In the context of change point detection, most of the transformations are unsupervised, for instance principal component analysis or linear discriminant analysis~\cite{Hastie2009}.
Supervised strategies are more rarely found, even though there exist numerous methods to learn a task-specific matrix $M$ in the context of supervised classification~\cite{Hastie2009,Xing2003,Davis2007}.
Those strategies fall under the umbrella of metric learning algorithms.
In the change point detection literature, there is only one work that proposes a supervised procedure to calibrate a metric matrix $M$~\cite{Lajugie2014}.
In this contribution, the authors use a training set of annotated signals (meaning that an expert has provided the change point locations) to learn $M$ iteratively.
Roughly, at each step, a new matrix $M$ is generated in order to improve change point detection accuracy on the training signals.
However, using the cost function $c_M$ is not adapted to certain applications, where a linear treatment of the data is insufficient.
In that situation, a well-chosen non-linear transformation of the data samples must be applied beforehand~\cite{Lajugie2014}.
%
%
\subsection{Non-parametric models}
When the assumptions of parametric models are not adapted to the data at hand, non-parametric change point detection methods can be more robust.
Three major approaches are presented here, each based on different non-parametric statistics, such as the empirical cumulative distribution function, rank statistics and kernel estimation.
\paragraph{Signal model.}
Assume that the observed signal $y=\{y_1,\dots,y_T\}$ is composed of independent random variables, such that
\begin{mymodel}\label{eq:cost_np_model}
	y_t \sim \sum_{k=0}^{K^{*}}\ F_k \ \one( t^{*}_k < t \leq t^{*}_{k+1} )
\end{mymodel}
where the $t^{*}_k$ are change point indexes and the $F_k$ are cumulative distribution functions (c.d.f.\!), not necessarily parametric as in~\eqref{eq:cost_pw_iid_model}.
Under this setting, the sub-signal $y_{t^{*}_k..t^{*}_{k+1}}$, bounded by two change points, is composed of iid variables with c.d.f.\ $F_k$.
When the $F_k$ belong to a known parametric distribution family, change point detection is performed with the MLE approach described in Section~\ref{sec:cost_iid}, which consists in applying the cost function $c_{\mbox{i.i.d.}}$.
However, this approach is not possible when the distribution family is either non-parametric or not known beforehand.
\subsubsection{Non-parametric maximum likelihood}
The first non-parametric cost function example, denoted $c_{\widehat{F}}$, has been introduced for the \emph{single} change point detection problem in~\cite{Einmahl2003} and extended for \emph{multiple} change points in~\cite{Zou2014}.
It relies on the empirical cumulative distribution function, estimated on sub-signals.
Formally, the signal is assumed to be univariate (ie $d=1$) and the empirical cdf on the sub-signal $y_{a..b}$ is given by
\begin{equation}
	\forall u\in\mathbb{R}, \quad\widehat{F}_{a..b} (u) := \frac{1}{b-a}\ \bigg[ \sum_{t=a+1}^b \one(y_t < u) + 0.5 \times\one(y_t=u)\bigg].
	\label{eq:review-emp-cdf}
\end{equation}
In order to derive a log-likelihood function that does not depend on the probability distribution of the data, ie the $f(\cdot|\theta_k)$, the authors use the following fact: for a fixed $u\in\mathbb{R}$, the empirical cdf $\widehat{F}$ of $n$ iid random variables, distributed from a certain cdf $F$ is such that $n\widehat{F}(u)\sim \mbox{Binomial}(n, F(u))$~\cite{Zou2014}.
This observation, combined with careful summation over $u$, allows a distribution-free maximum likelihood estimation.
The resulting cost function $c_{\widehat{F}}$ is defined as follows.
Interestingly, this strategy was first introduced to design non-parametric two-sample statistical tests, which were experimentally shown to be more powerful than classical tests such as Kolmogorov-Smirnov and Cramér-von Mises~\cite{Einmahl2003, Zhang2006}.
\begin{cost}[$c_{\widehat{F}}$]
	\label{def:review-cdf}
	The cost function $c_{\widehat{F}}$ is given by
	\begin{mycost}
		c_{\widehat{F}}(y_{a..b}) := - (b-a)\ \sum_{u=1}^{T} \frac{\widehat{F}_{a..b}(u)\log\widehat{F}_{a..b}(u) + (1-\widehat{F}_{a..b}(u))\log(1-\widehat{F}_{a..b}(u))}{(u-0.5)(T-u+0.5)}
		\label{eq:review-c-cdf}
	\end{mycost}
	where the empirical cdf $\widehat{F}_{a..b}$ is defined by~\eqref{eq:review-emp-cdf}.
\end{cost}
\noindent
%
%
From a theoretical point of view, asymptotic consistency of change point estimates is verified, when the number of change points is either known or unknown~\cite{Zou2014}.
However, solving either one of the detection problems can be computationally intensive, because calculating the value of the cost function $c_{\widehat{F}}$ on one sub-signal requires to sum $T$ terms, where $T$ is the signal length.
As a result, the total complexity of change point detection is of the order of $\mathcal{O}(T^3)$~\cite{Zou2014}.
To cope with this computational burden, several preliminary steps are proposed.
For instance, irrelevant change point indexes can be removed before performing the detection, thanks to a screening step~\cite{Zou2014}.
Also, the cost function $c_{\widehat{F}}$ can be approximated, by summing, in~\eqref{eq:review-c-cdf}, over a few (carefully chosen) terms, instead of $T$ terms originally~\cite{Haynes2017a}.
Thanks to those implementation techniques, the cost function $c_{\widehat{F}}$ has been applied on DNA sequences~\cite{Zou2014} and heart-rate monitoring signals~\cite{Haynes2017a}.
\subsubsection{Rank-based detection}
In statistical inference, a popular strategy to derive distribution-free statistics is to replaced the data samples by their ranks within the set of pooled observations~\cite{Levy-Leduc2009, Clemencon2009,Friedman1979}.
In the context of change point detection, this strategy has first been applied to detect a \emph{single} change point~\cite{Levy-Leduc2009, Lung-Yut-Fong2012}, and then has been extended by~\cite{Lung-Yut-Fong2015} to find \emph{multiple} change points.
The associated cost function, denoted $c_{\mbox{rank}}$, is defined as follows.
Formally, it relies on the centered $\mathbb{R}^d$-valued ``rank signal'' $r=\{r_t\}_{t=1}^T$, given by
\begin{equation}
	r_{t, j} := \sum_{s=1}^T \one(y_{s,j} \leq y_{t,j})\ - \ \frac{T+1}{2} ,\quad \forall 1\leq t\leq T, \ \forall 1\leq j \leq d.
	\label{eq:review-rank-signal}
\end{equation}
In other words, $r_{t,j}$ is the (centered) rank of the $j^{\mbox{th}}$ coordinate of the $t^{\mbox{th}}$ sample, ie $y_{t, j}$, among the $ \{y_{1, j},y_{2, j},\dots,y_{T, j}\}$.
\begin{cost}[$c_{\mbox{rank}}$]
	The cost function $c_{\mbox{rank}}$ is given by
	\begin{mycost}
		c_{\mbox{rank}}(y_{a..b}) := - (b-a) \ \bar{r}_{a..b}' \ \widehat{\Sigma}_r^{-1}\ \bar{r}_{a..b}
		\label{eq:review-cost-rank}
	\end{mycost}
	where the signal $r$ is defined in~\eqref{eq:review-rank-signal} and $\widehat{\Sigma}_r\in\mathbb{R}^{d \times d}$ is the following matrix
	\begin{equation}
		\widehat{\Sigma}_r  := \frac{1}{T} \sum_{t=1}^T (r_t + 1/2)' (r_t + 1/2).
		\label{eq:review-rank-covariance}
	\end{equation}
\end{cost}
\noindent
Intuitively, $c_{\mbox{rank}}$ measures changes in the joint behaviour of the marginal rank statistics of each coordinate, which are contained in $r$.
One of the advantages of this cost function is that it is invariant under any monotonic transformation of the data.
Several well-known statistical hypothesis testing procedures are based on this scheme, for instance the Wilcoxon-Mann-Whitney test~\cite{Wilcoxon1945}, the Friedman test~\cite{Neath2006}, the Kruskal-Wallis test~\cite{Kendall1970}, and several others~\cite{Clemencon2009,Friedman1979}.
From a computational point of view, two steps must be performed before the change point detection: the calculation of the rank statistics, in $\mathcal{O}(dT\log T)$ operations, and the calculation of the matrix $\widehat{\Sigma}_r$, in $\mathcal{O}(d^2 T + d^3)$ operations.
The resulting algorithm has been applied on DNA sequences~\cite{Lung-Yut-Fong2015} and network traffic data~\cite{Levy-Leduc2009, Lung-Yut-Fong2012}.
\subsubsection{Kernel-based detection}\label{sec:cost_kernel}
A kernel-based method has been proposed by~\cite{harchaoui2007retrospective} to perform change point detection in a non-parametric setting.
To that end, the original signal $y$ is mapped onto a reproducing Hilbert space (rkhs) $\mathcal{H}$ associated with a user-defined kernel function $k(\cdot, \cdot): \mathbb{R}^d \times \mathbb{R}^d \to \mathbb{R} $.
The mapping function $\phi:\mathbb{R}^d\to\mathcal{H}$ onto this rkhs is implicitly defined by $\phi(y_t) = k(y_t,\cdot) \in \mathcal{H}$, resulting in the following inner-product and norm:
\begin{equation}
	\sca{\phi(y_s)}{\phi(y_t)}_\mathcal{H} = k(y_s,y_t) \quad\mbox{and}\quad \norm{\phi(y_t)}_\mathcal{H}^2 = k(y_t,y_t)
	\label{eq:review-kernel-implicit}
\end{equation}
for any samples $y_s,y_t\in\mathbb{R}^d$.
The associated cost function, denoted $c_{\mbox{kernel}}$, is defined as follows.
This kernel-based mapping is central to many machine learning developments such as support vector machine or clustering~\cite{Scholkopf2002,gretton2012kernel}.

\begin{cost}[$c_{\mbox{kernel}}$]
	\label{def:review-cost-kernel}
	For a given kernel function $k(\cdot, \cdot): \mathbb{R}^d \times \mathbb{R}^d \to \mathbb{R} $, the cost function $c_{\mbox{kernel}}$ is given by
	\begin{mycost}
		c_{\mbox{kernel}}(y_{a..b}) := \sum_{t=a+1}^{b} \norm{\phi(y_t) - \bar{\mu}_{a..b}}_{\mathcal{H}}^2
		\label{eq:cost_kernel}
	\end{mycost}
	where $\bar{\mu}_{a..b}\in\mathcal{H}$ is the empirical mean of the embedded signal $\{\phi(y_t)\}_{t=a+1}^{b}$ and $\norm{\cdot}_\mathcal{H}$ is defined in~\eqref{eq:review-kernel-implicit}.
\end{cost}

\begin{remark}[Computing the cost function]
	Thanks to the well-known ``kernel trick'', the explicit computation of the mapped data samples $\phi(y_t)$ is not required to calculate the cost function value~\cite{Celisse2017}.
	Indeed, after simple algebraic manipulations, $c_{\mbox{kernel}}(y_{a..b})$ can be rewritten as follows:
	\begin{equation}
		c_{\mbox{kernel}}(y_{a..b}) = \sum_{t=a+1}^{b} k(y_t,y_t) \ -\ \frac{1}{b-a} \sum_{s,t=a+1}^{b} k(y_s,y_t).
	\end{equation}
\end{remark}

\begin{remark}[Intuition behind the cost function]
	\label{rem:cost_mean_embedding}
	Intuitively, the cost function $c_{\mbox{kernel}}$ is able to detect mean-shifts in the transformed signal $\{\phi(y_t)\}_t $.
	Its use is motivated in the context of Model~\ref{eq:cost_np_model} by the fact that, under certain conditions on the kernel function, changes in the probability distribution coincide with mean-shifts in the transformed signal.
	This connection has been investigated in several works on kernel methods~\cite{sriperumbudur2008injective,Scholkopf2002,Shawe-Taylor2004,gretton2012kernel}.
	Formally, let $\mathbb{P}$ denote a probability distribution defined over $\mathbb{R}^d$.
	Then there exists a unique element $\mu_\mathbb{P}\in\mathcal{H}$~\cite{sriperumbudur2008injective}, called the mean embedding (of $\mathbb{P}$), such that
	\begin{equation}
		\mu_\mathbb{P} = \mathbb{E}_{X\sim\mathbb{P}}\ [\phi(X)].
		\label{eq:cost_mean_embedding}
	\end{equation}
	In addition, the mapping $\mathbb{P}\mapsto\mu_\mathbb{P}$ is injective (in which case the kernel is said to be \emph{characteristic}), meaning that
	\begin{equation}
		\mu_\mathbb{P} = \mu_\mathbb{Q} \Longleftrightarrow \mathbb{P}=\mathbb{Q},
	\end{equation}
	where $\mathbb{Q}$ denotes a probability distribution defined over $\mathbb{R}^d$.
	In order to determine if a kernel is characteristic (and therefore, useful for change point detection), several conditions can be found in the literature~\cite{sriperumbudur2008injective, Scholkopf2002,gretton2012kernel}.
	For instance, if a kernel $k(\cdot,\cdot)$ is translation invariant, meaning that $k(y_s,y_t)=\psi(y_s-y_t)\ \forall s,t$, where $\psi$ is a bounded continuous positive definite function on $\mathbb{R}^d$, then it is characteristic~\cite{sriperumbudur2008injective}.
	This condition is verified by the commonly used Gaussian kernel.
	As a consequence, two transformed samples $\phi(y_s)$ and $\phi(y_t)$ are distributed around the \emph{same} mean value if they belong to the same regime, and around \emph{different} mean-values if they each belong to two consecutive regimes.
	To put it another way, a signal that follows~\eqref{eq:cost_np_model} is mapped by $\phi(\cdot)$ to a random signal with piecewise constant mean.
\end{remark}%
\noindent
From a theoretical point of view, asymptotic consistency of the change point estimates has been demonstrated for both a known and unknown number of change points in the recent work of \cite{Garreau2016}.
This result, as well as an important oracle inequality on the sum of cost $V(\mathcal{T})$~\cite{arlot2012kernel}, also holds in a non-asymptotic setting.
In addition, kernel change point detection was experimentally shown to be competitive in many different settings, in an unsupervised manner and with very few parameters to manually calibrate.
For instance, the cost function $c_{\mbox{kernel}}$ was applied on the Brain-Computer Interface (BCI) data set~\cite{harchaoui2007retrospective}, on a video time series segmentation task~\cite{arlot2012kernel}, DNA sequences~\cite{Celisse2017} and emotion recognition~\cite{Cabrieto2018}.
\paragraph{Related cost functions.}
The cost function $c_{\mbox{kernel}}$ can be combined with any kernel to accommodate various types of data (not just $\mathbb{R}^d$-valued signals).
Notable examples of kernel functions include~\cite{Shawe-Taylor2004}:
\begin{itemize}
	\item The linear kernel $k(x,y) = \sca{x}{y}$ with $x,y \in\mathbb{R}^d$.
	\item The polynomial kernel $k(x,y) = (\sca{x}{y} + C)^{deg}$ with $x,y \in\mathbb{R}^d$, and $C$ and $deg$ are parameters.
	\item The Gaussian kernel $k(x,y)=\exp(-\gamma\norm{x-y}^2)$ with $x,y\in\mathbb{R}^d$ and $\gamma>0$ is the so-called bandwidth parameter.
	\item The $\chi^2$-kernel $k(x, y) = \exp(-\gamma \sum_i [(x_i - y_i)^2 / (x_i + y_i)])$ with $\gamma\in\mathbb{R}$ a parameter.
	      It is often used for histogram data.
\end{itemize}%
Arguably, the most commonly used kernels for numerical data are the linear kernel and the Gaussian kernel.
When combined with the linear kernel, the cost function $c_{\mbox{kernel}}$ is formally equivalent to $c_{L_2}$.
As for the Gaussian kernel, the associated cost function, denoted $c_{\mbox{rbf}}$, is defined as follows.
\begin{cost}[$c_{\mbox{rbf}}$]
	The cost function $c_{\mbox{rbf}}$ is given by
	\begin{mycost}
		c_{\mbox{rbf}}(y_{a..b}) := (b-a)\ -\ \frac{1}{b-a} \sum_{s,t=a+1}^{b} \exp(-\gamma\norm{y_s-y_t}^2)
		\label{eq:cost_rbf}
	\end{mycost}
	where $\gamma>0$ is the so-called bandwidth parameter.
\end{cost}
\noindent
The parametric cost function $c_M$ (based on a Mahalanobis-type norm) can be extended to the non-parametric setting through the use of a kernel.
Formally, the Mahalanobis-type norm $\norm{\cdot}_{\mathcal{H},M}$ in the feature space $\mathcal{H}$ is defined by
\begin{equation}
	\norm{\phi(y_s)-\phi(y_t)}^2_{\mathcal{H},M} = (\phi(y_s)-\phi(y_t))'\,M\,(\phi(y_s)-\phi(y_t))
	\label{eq:review-maha-kernel}
\end{equation}
where $M$ is a (possibly infinite dimensional) symmetric positive semi-definite matrix defined on $\mathcal{H}$.
The associated cost function, denoted $c_{\mathcal{H},M}$, is defined below.
Intuitively, using $c_{\mathcal{H},M}$ instead of $c_M$ introduces a non-linear treatment of the data samples.
\begin{cost}[$c_{\mathcal{H},M}$]
	For a given kernel function $k(\cdot, \cdot): \mathbb{R}^d \times \mathbb{R}^d \to \mathbb{R} $ and $M$ a symmetric positive semi-definite matrix defined on the associated rkhs $\mathcal{H}$, the cost function $c_{\mathcal{H},M}$ is given by
	\begin{mycost}
		c_{\mathcal{H},M}(y_{a..b}) := \sum_{t=a+1}^{b} \norm{\phi(y_t) - \bar{\mu}_{a..b}}^2_{\mathcal{H},M}
		\label{eq:cost_kml}
	\end{mycost}
	where $\mu_{a..b}$ is the empirical mean of the transformed sub-signal $\{\phi(y_t)\}_{t=a+1}^b$ and $\norm{\cdot}_{\mathcal{H},M}$ is defined in~\eqref{eq:review-maha-kernel}.
\end{cost}
%
%
\subsection{Summary table}
Reviewed cost functions (parametric and non-parametric) are summarized in Table~\ref{tab:cost_summary}.
For each cost, the name, expression and parameters of interest are given.

\begin{table}[H]
	\centering
	\scriptsize
	\begin{tabular}{clp{15em}}
		Name                                                   & $c(y_{a..b})$                                                                                                                                              & Parameters                                                                                                                          \\ \hline \\
		$c_{\mbox{i.i.d.}}$~\eqref{eq:cost_general_likelihood} & $- \sup_{\theta} \sum_{t=a+1}^{b} \log f(y_t|\theta)$                                                                                                      & $\theta$: changing parameter; density function: $f(\cdot|\theta)$                                                                   \\[1em]
		$c_{L_2}$~\eqref{eq:cost_mean_shift}                   & $ \sum_{t=a+1}^{b} \norm{y_t - \bar{y}_{a..b}}^2_2$                                                                                                        & $\bar{y}_{a..b}$: empirical mean of $y_{a..b}$                                                                                      \\[1em]
		$c_{\Sigma}$~\eqref{eq:cost_scale_shift}               & $(b-a) \log\det\widehat{\Sigma}_{a..b} + \sum_{t=a+1}^b (y_t - \bar{y}_{a..b})' \widehat{\Sigma}_{a..b}^{-1} (y_t - \bar{y}_{a..b}) $                      & $\widehat{\Sigma}_{a..b}$: empirical covariance of $y_{a..b}$                                                                       \\[1em]
		$c_{\mbox{Poisson}}$~\eqref{eq:cost_poisson}           & $ - (b-a) \bar{y}_{a..b}\log\bar{y}_{a..b}$                                                                                                                & $\bar{y}_{a..b}$: empirical mean of $y_{a..b}$                                                                                      \\[1em]
		$c_{\mbox{linear}}$~\eqref{eq:cost_linear}             & $ \min_{u\in\mathbb{R}^p, v\in\mathbb{R}^q}\ \sum_{t=a+1}^{b} (y_t - x_t'u - z_t'v)^2$                                                                                   & $x_t\in\mathbb{R}^p,z_t\in\mathbb{R}^q$: covariates                                                                                               \\[1em]
		$c_{\mbox{linear}, L_1}$~\eqref{eq:linear_cost_abs}    & $ \min_{u\in\mathbb{R}^p, v\in\mathbb{R}^q}\ \sum_{t=a+1}^{b} |y_t - x_t'u - z_t'v|$                                                                                     & $x_t\in\mathbb{R}^p,z_t\in\mathbb{R}^q$: covariates                                                                                               \\[1em]
		$c_{\mbox{AR}}$~\eqref{eq:linear_cost_ar}              & $ \min_{u\in\mathbb{R}^p}\ \sum_{t=a+1}^{b} (y_t - x_t'u)^2$                                                                                                      & $x_t=[y_{t-1},y_{t-2},\dots,y_{t-p}]$: lagged samples                                                                               \\[1em]
		$c_M$~\eqref{eq:cost_mahalanobis}                      & $\sum_{t=a+1}^{b} \norm{y_t - \bar{y}_{a..b}}^2_M$                                                                                                         & $M\in\mathbb{R}^{d\times d}$: positive semi-definite matrix                                                                                \\[1em]
		$c_{\widehat{F}}$~\eqref{eq:review-c-cdf}              & $- (b-a)\ \sum_{u=1}^{T} \frac{\widehat{F}_{a..b}(u)\log\widehat{F}_{a..b}(u) + (1-\widehat{F}_{a..b}(u))\log(1-\widehat{F}_{a..b}(u))}{(u-0.5)(T-u+0.5)}$ & $\widehat{F}$: empirical c.d.f.~\eqref{eq:review-emp-cdf}                                                                           \\[1em]
		$c_{\mbox{rank}}$~\eqref{eq:review-cost-rank}          & $- (b-a) \ \bar{r}_{a..b}' \ \widehat{\Sigma}_r^{-1}\ \bar{r}_{a..b}$                                                                                      & $r$: rank signal~\eqref{eq:review-rank-signal}; $\widehat{\Sigma}_r$: empirical covariance of $r$~\eqref{eq:review-rank-covariance} \\[1em]
		$c_{\mbox{kernel}}$~\eqref{eq:cost_kernel}             & $ \sum_{t=a+1}^{b} k(y_t,y_t) \ -\ \frac{1}{b-a} \sum_{s,t=a+1}^{b} k(y_s,y_t)$                                                                            & $k(\cdot,\cdot):\mathbb{R}^d \times \mathbb{R}^d \mapsto\mathbb{R}$: kernel function                                                                     \\[1em]
		$c_{\mbox{rbf}}$~\eqref{eq:cost_rbf}                   & $ (b-a)\ -\ \frac{1}{b-a} \sum_{s,t=a+1}^{b} \exp(-\gamma\norm{y_s-y_t}^2)$                                                                                & $\gamma>0$: bandwidth parameter                                                                                                     \\[1em]
		$c_{\mathcal{H},M}$~\eqref{eq:cost_kml}                        & $\sum_{t=a+1}^{b} \norm{y_t - \bar{y}_{a..b}}^2_{\mathcal{H},M}$                                                                                                   & $M$: positive semi-definite matrix (in the feature space $\mathcal{H}$)                                                                     
	\end{tabular}
	\caption{Summary of cost reviewed functions}
	\label{tab:cost_summary}
\end{table}

\section{Search methods}\label{sec:review_kn}
\begin{figure}[t]
	\centering
	\includegraphics[width=0.8\columnwidth]{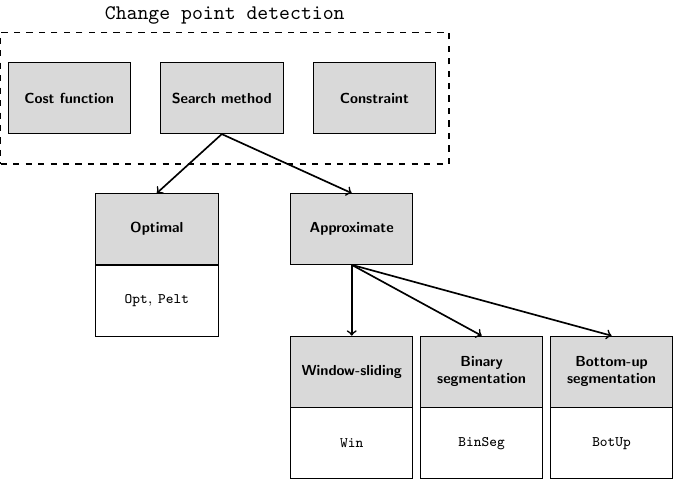}
	\caption{Typology of the search methods described in Section~\ref{sec:review_kn}.}
	\label{fig:review-diagram-search}
\end{figure}

This section presents the second defining element of change detection methods, namely the search method.
Reviewed search methods are organized in two general categories, as shown on Figure~\ref{fig:review-diagram-search}: optimal methods, that yield the exact solution to the discrete optimization of \eqref{eq:review_kn} and \eqref{eq:review_unkn}, and the approximate methods, that yield an approximate solution.
Described algorithms can be combined with cost functions from Section~\ref{sec:review_costs}.
Note that, depending on the chosen cost function, the computational complexity of the complete algorithm changes.
As a consequence, in the following, complexity analysis is done with the assumption that applying the cost function on a sub-signal requires $\mathcal{O}(1)$ operations.
Also, the practical implementations of the most important algorithms are given in pseudo-code.

\subsection{Optimal detection}

Optimal detection methods find the exact solutions of Problem 1 \eqref{eq:review_kn} and Problem 2 \eqref{eq:review_unkn}.
A naive approach consists in enumerating all possible segmentations of a signal, and returning the one that minimizes the objective function.
However, for \eqref{eq:review_kn}, minimization is carried out over the set $\{\mathcal{T}\ \mbox{s.t.}\ |\mathcal{T}|=K\} $ (which contains ${\binom{T-1}{K-1}}$ elements), and for \eqref{eq:review_unkn}, over the set $\{\mathcal{T}\ \mbox{s.t.}\ 1\leq|\mathcal{T}|<T\} $ (which contains $\sum_{K=1}^{T-1}  {\binom{T-1}{K-1}}$ elements).
This makes exhaustive enumeration impractical, in both situations.
We describe in this section two major approaches to efficiently find the exact solutions of \eqref{eq:review_kn} and \eqref{eq:review_unkn}.

\subsubsection{Solution to Problem 1 \eqref{eq:review_kn}: \texttt{Opt}}
In \eqref{eq:review_kn}, the number of change points to detect is fixed to a certain $K\geq1$.
The optimal solution to this problem can be computed efficiently, thanks to a method based on dynamic programming.
The algorithm, denoted \texttt{Opt}, relies on the additive nature of the objective function $V(\cdot)$ to recursively solve sub-problems.
Precisely, \texttt{Opt} is based on the following observation:
\begin{equation}
	\begin{split}
		\min_{|\mathcal{T}|=K}\ V(\mathcal{T}, y=y_{0..T} ) &= \min_{0=t_0<t_1<\dots<t_K<t_{K+1}=T}\ \sum_{k=0}^{K} \ c(y_{t_{k}..t_{k+1}}) \\
		&= \min_{t\leq T-K} \ \bigg[ c(y_{0..t})\ +\ \min_{t=t_0<t_1<\dots<t_{K-1}<t_{K}=T} \sum_{k=0}^{K-1} \ c(y_{t_{k}..t_{k+1}}) \bigg] \\
		&= \min_{t\leq T-K} \ \bigg[ c(y_{0..t})\ +\ \min_{|\mathcal{T}|=K-1} V(\mathcal{T}, y_{t..T} ) \bigg]
	\end{split}
	\label{eq:opt_sub_problem}
\end{equation}
Intuitively, Equation~\ref{eq:opt_sub_problem} means that the first change point of the optimal segmentation is easily computed if the optimal partitions with $K-1$ elements of all sub-signals $y_{t..T} $ are known.
The complete segmentation is then computed by recursively applying this observation.
This strategy, described in detail in Algorithm~\ref{alg:opt}, has a complexity of the order of $\mathcal{O}(KT^2)$~\cite{Kay1993, bai2003computation}.
Historically, \texttt{Opt} was introduced for a non-related problem~\cite{bellman1956routing} and later applied to change point detection, in many different contexts, such as EEG recordings~\cite{Lavielle2005,Lavielle1998}, DNA sequences~\cite{Rigaill,Celisse2017}, tree growth monitoring~\cite{guedon:inria-00311634}, financial time-series~\cite{Lavielle1999,Perron2006}, radar waveforms~\cite{Rossi2009simultaneousclustering}, etc.
\begin{algorithm}[H]
	\caption{Algorithm \texttt{Opt}}
	\label{alg:opt}
	\begin{algorithmic}
		\State\textbf{Input:} signal $\{y_t\}_{t=1}^{T}$, cost function $c(\cdot)$, number of regimes $K\geq2$.
		\ForAll{$(u,v),\ 1\leq u<v \leq T$}
		\State Initialize $C_1 (u,v) \gets c(\{y_t\}_{t=u}^{v}).$
		\EndFor
		\For{$k=2,\dots,K-1$}
		\ForAll{$u,v\in\{1,\dots,T\}, v-u\geq k$}
		\State $C_k (u,v) \gets \min\limits_{u+k-1\leq t<v}  C_{k-1}(u,t) + C_{1}(t+1,v)$
		\EndFor
		\EndFor
		\State Initialize $L$, a list with $K$ elements.
		\State Initialize the last element: $L[K] \gets T$.
		\State Initialize $k\gets K$.
		\While{$k>1$}
		\State $s\gets L(k)$
		\State $t^{*} \gets \operatorname{argmin}\limits_{k-1\leq t<s} C_{k-1}(1, t) + C_1(t+1, s)$
		\State $L(k-1) \gets t^{*}$
		\State $k\gets k-1$
		\EndWhile
		\State Remove $T$ from $L$
		\State\textbf{Output:} set $L$ of estimated breakpoint indexes.
	\end{algorithmic}
\end{algorithm}
\paragraph{Related search methods.}
Several extensions of \texttt{Opt} have been proposed in the literature.
The proposed methods still find the exact solution to \eqref{eq:review_kn}.
\begin{itemize}
	\item[-] The first extension is the ``forward dynamic programming'' algorithm~\cite{guedon:inria-00311634}.
	      Contrary to \texttt{Opt}, which returns a single partition, the ``forward dynamic programming'' algorithm computes the top $L$ ($L\geq 1$) most probable partitions (ie with lowest sum of costs).
	      The resulting computational complexity is $\mathcal{O}(LKT^2)$ where $L$ is the number of computed partitions.
	      This method is designed as a diagnostic tool: change points present in many of the top partitions are considered very likely, while change points present in only a few of the top partitions might not be as relevant.
	      Thanks to ``forward dynamic programming'', insignificant change points are trimmed and overestimation of the number of change point is corrected~\cite{guedon:inria-00311634}, at the expense of a higher computational burden.
	      It is applied on tree growth monitoring time series~\cite{guedon:inria-00311634} that are relatively short with around a hundred samples.
	\item[-] The ``pruned optimal dynamic programming'' procedure~\cite{Rigaill} is an extension of \texttt{Opt} that relies on a pruning rule to discard indexes that can never be change points.
	      Thanks to this trick, the set of potential change point indexes is reduced.
	      All described cost functions can be plugged into this method.
	      As a result, longer signals can be handled, for instance long array-based DNA copy number data (up to $10^6$ samples, with the quadratic error cost function)~\cite{Rigaill}.
	      However, worst case complexity remains of the order of $\mathcal{O}(KT^2)$.
\end{itemize}
\subsubsection{Solution to Problem~2~\eqref{eq:review_unkn}: \texttt{Pelt}}\label{sec:review-pelt}
In \eqref{eq:review_unkn}, the number of changes point is unknown, and the objective function to minimize is the penalized sum of costs.
A naive approach consists in applying \texttt{Opt} for $K=1,\dots,K_{\mbox{max}}$ for a sufficiently large $K_{\mbox{max}}$, then choosing among the computed segmentations the one that minimizes the penalized problem.
This would prove computational cumbersome because of the quadratic complexity of the resolution method \texttt{Opt}.
Fortunately a faster method exists for a general class of penalty functions, namely linear penalties.
Formally, linear penalties are linear functions of the number of change points, meaning that
\begin{equation}
	\mbox{pen}(\mathcal{T}) = \beta |\mathcal{T}|
	\label{eq:review-linear-pen}
\end{equation}
where $\beta>0$ is a smoothing parameter.
(More details on such penalties can be found in Section~\ref{sec:review-linear-pen}.)
The algorithm \texttt{Pelt} (for ``Pruned Exact Linear Time'')~\cite{Killick2012a} was introduced to find the exact solution of \eqref{eq:review_unkn}, when the penalty is linear~\eqref{eq:review-linear-pen}.
This approach considers each sample sequentially and, thanks to an explicit pruning rule, may or may not discard it from the set of potential change points.
Precisely, for two indexes $t$ and $s$ ($t<s<T$), the pruning rule is given by:
\begin{multline}
	\mbox{if}\quad	\bigg[ \min_{\mathcal{T}} V(\mathcal{T}, y_{0..t}) + \beta |\mathcal{T}| \bigg] + c(y_{t..s}) \geq \bigg[ \min_{\mathcal{T}} V(\mathcal{T}, y_{0..s}) + \beta |\mathcal{T}| \bigg] \quad\mbox{holds,}\\
	\mbox{then } t \mbox{ cannot be the last change point prior to } T.
\end{multline}
This results in a considerable speed-up: under the assumption that regime lengths are randomly drawn from a uniform distribution, the complexity of \texttt{Pelt} is of the order $\mathcal{O}(T)$.
The detailed algorithm can be found in Algorithm~\ref{alg:pelt}.
An extension of \texttt{Pelt} is described in~\cite{Haynes2017} to solve the linearly penalized change point detection for a range of smoothing parameter values $[\beta_{\mbox{min}},\beta_{\mbox{max}}]$.
\texttt{Pelt} has been applied on DNA sequences~\cite{Hocking2013, Maidstone2017}, physiological signals~\cite{Haynes2017a}, and oceanographic data~\cite{Killick2012a}.
\begin{algorithm}[H]
	\caption{Algorithm \texttt{Pelt}}
	\begin{algorithmic}
		\State\textbf{Input:} signal $\{y_t\}_{t=1}^{T}$, cost function $c(\cdot)$, penalty value $\beta$.
		\State Initialize $Z$ a $(T+1)$-long array; $Z[0]\gets -\beta$.
		\State Initialize $L[0] \gets \emptyset $.
		\State Initialize $\chi \gets \{0\}$. \Comment{Admissible indexes.}
		\For{$t=1,\dots,T$}
		\State $\hat{t} \gets \operatorname{argmin}_{s\in \chi} \big[Z[s] + c(y_{s..t}) + \beta\big]$.
		\State $Z[t] \gets \big[Z[\hat{t}] + c(y_{\hat{t}..t}) + \beta\big]$
		\State $L[t] \gets L[\hat{t}] \cup \{\hat{t}\} $.
		\State $\chi \gets \{ s \in \chi \; : \; Z[s] + c(y_{s..t}) \leq Z[t]\} \cup \{t\} $
		\EndFor
		\State\textbf{Output:} set $L[T]$ of estimated breakpoint indexes.
	\end{algorithmic}
	\label{alg:pelt}
\end{algorithm}
%

\subsection{Approximate detection}
When the computational complexity of optimal methods is too great for the application at hand, one can resort to approximate methods.
In this section, we describe three major types of approximate segmentation algorithms, namely window-based methods, binary segmentation and bottom-up segmentation.
All described procedures fall into the category of sequential detection approaches, meaning that they return a single change point estimate $\hat{t}^{(k)}$ ($1\leq \hat{t}^{(k)} <T $) at the $k$-th iteration.
(In the following, the subscript $\cdot^{(k)}$ refers to the $k$-th iteration of a sequential algorithm.)
Such methods can be used to solve (approximately) either \eqref{eq:review_kn} or \eqref{eq:review_unkn}.
Indeed, if the number $K^{*}$ of changes is known, $K^{*}$ iterations of a sequential algorithm are enough to retrieve a segmentation with the correct number of changes.
If $K^{*}$ is unknown, the sequential algorithm is run until an appropriate stopping criterion is met.

%
%
\subsubsection{Window sliding}
\label{sec:review_win}
\begin{figure}[t]
	\centering
	\includegraphics[width=\columnwidth]{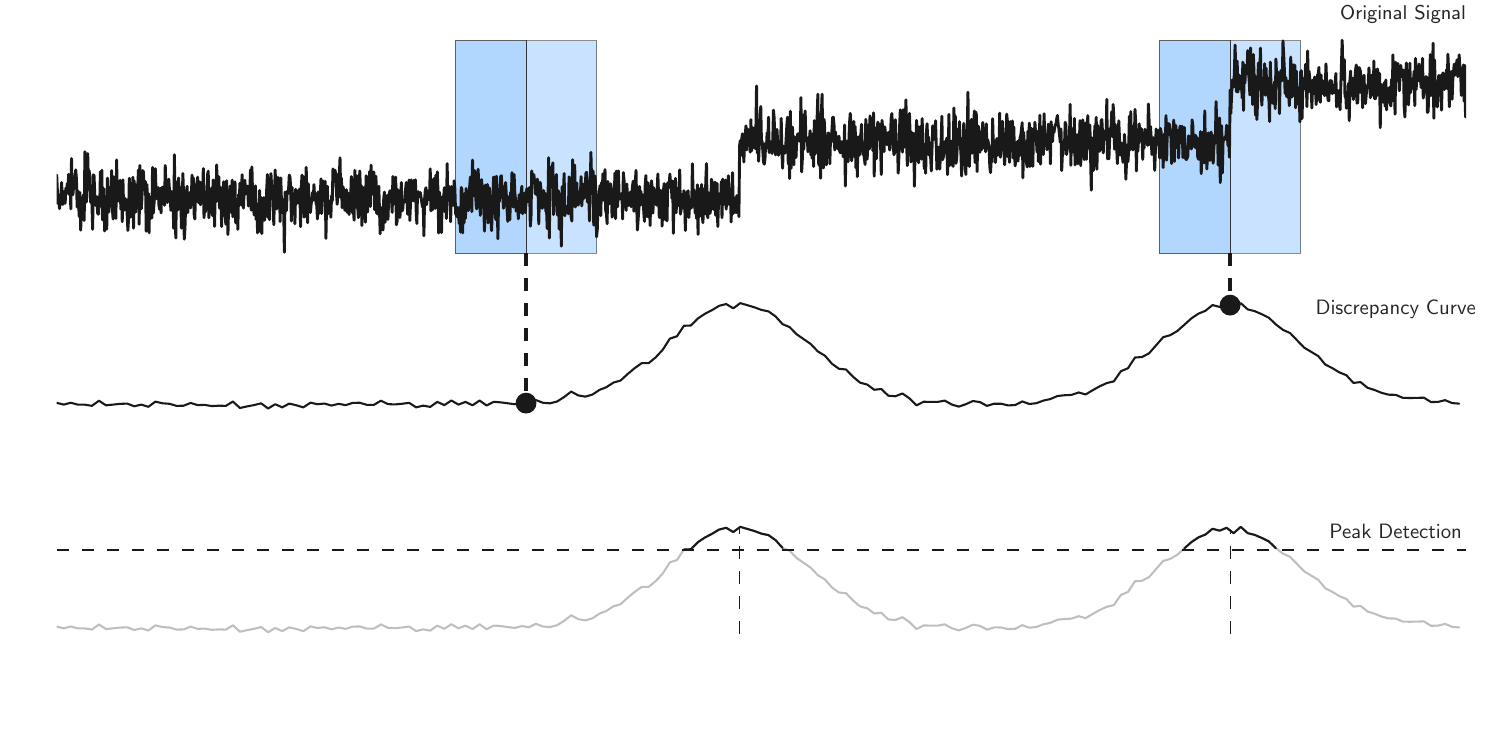}
	\caption{Schematic view of \texttt{Win}}
	\label{fig:win_schema}
\end{figure}

The window-sliding algorithm, denoted \texttt{Win}, is a fast approximate alternative to optimal methods.
It consists in computing the discrepancy between two adjacent windows that slide along the signal $y$.
For a given cost function $c(\cdot)$, this discrepancy between two sub-signals is given by
\begin{equation}
	d (y_{a..t}, y_{t..b}) = c(y_{a..b}) - c(y_{a..t}) - c(y_{t..b}) \quad (1\leq a < t < b \leq T).
	\label{eq:win_discrepancy}
\end{equation}%
When the two windows cover dissimilar segments, the discrepancy reaches large values, resulting in a peak.
In other other words, for each index $t$, \texttt{Win} measures the discrepancy between the immediate past (``left window'') and the immediate future (``right window'').
Once the complete discrepancy curve has been computed, a peak search procedure is performed to find change point indexes.
The complete \texttt{Win} algorithm is given in Algorithm~\ref{alg:win} and a schematic view is displayed on Figure~\ref{fig:win_schema}.
The main benefits of \texttt{Win} are its low complexity (linear in the number of samples) and ease of implementation.
\begin{algorithm}[H]
	\caption{Algorithm \texttt{Win}}
	\begin{algorithmic}
		\State\textbf{Input:} signal $\{y_t\}_{t=1}^{T}$, cost function $c(\cdot)$, half-window width $w$, peak search procedure {\tt PKSearch}.
		\State Initialize $Z\gets [0,0,\dots]$ a $T$-long array filled with $0$. \Comment{Score list.}
		\For{$t=w,\dots,T-w$}
		\State $p\gets (t-w)..t$.
		\State $q\gets t..(t+w)$.
		\State $r\gets (t-w)..(t+w)$.
		\State $Z[t] \gets c(y_{r}) - [c(y_{p}) + c(y_{q})] $.
		\EndFor
		\State $L\gets \mbox{\tt PKSearch}(Z)$ \Comment{Peak search procedure.}
		\State\textbf{Output:} set $L$ of estimated breakpoint indexes.
	\end{algorithmic}
	\label{alg:win}
\end{algorithm}
\noindent
In the literature, the discrepancy measure $d(\cdot,\cdot)$ is often derived from a two-sample statistical test (see Remark~\ref{rem:win_two_sample_test}), and not from a cost function, as in~\eqref{eq:win_discrepancy}.
However, the two standpoints are generally equivalent: for instance, using $c_{L_2}$, $c_{\mbox{i.i.d.}}$ or $c_{\mbox{kernel}}$ is respectively equivalent to applying a Student t-test~\cite{Basseville1993}, a generalized likelihood ratio (GLR)~\cite{Chen2011} test and a kernel Maximum Mean Discrepancy (MMD) test~\cite{gretton2012kernel}.
As a consequence, practitioners can capitalize on the vast body of work in the field of statistical tests to obtain asymptotic distributions for the discrepancy measure~\cite{gretton2012kernel,Chen1997,Lung-Yut-Fong2012,Levy-Leduc2009}, and sensible calibration strategies for important parameters of \texttt{Win} (such as the window size or the peak search procedure).
\texttt{Win} has been applied in numerous contexts: for instance, on biological signals~\cite{vullings1997ecg,E.Brodsky1999,Harchaoui2009,Esteller2001,Karagiannaki2017}, on network data~\cite{Levy-Leduc2009,Lung-Yut-Fong2012}, on speech time series~\cite{adak1998time,desobry2005online,Harchaoui2009} and on financial time series~\cite{chen1998speaker,Basseville1993,keogh2004segmenting}.
It should be noted that certain window-based detection methods in the literature rely on a discrepancy measure which is not related to a cost function, as in~\eqref{eq:win_discrepancy}~\cite{Harchaoui2009,harchaoui2008nips,Liu2013,Kifer:2004:DCD:1316689.1316707}.
As a result, those methods, initially introduced in the online detection setting, cannot be extended to work with optimal algorithms (\texttt{Opt}, \texttt{Pelt}).
\begin{remark}[Two-sample test]
	\label{rem:win_two_sample_test}
	A two-sample test (or homogeneity test) is a statistical hypothesis testing procedure designed to assess whether two populations of samples are identical in distribution.
	Formally, consider two sets of iid $\mathbb{R}^d$-valued random samples $\{x_t\}_t$ and $\{z_t\}_t$.
	Denote by $\mathbb{P}_x$ the distribution function of the $x_t$ and by $\mathbb{P}_z$, the distribution function of the $z_t$.
	A two-sample test procedure compares the two following hypotheses:
	\begin{equation}
		\begin{split}
			H_0 &:\quad \mathbb{P}_x = \mathbb{P}_z \\
			H_1 &:\quad \mathbb{P}_x \neq \mathbb{P}_z.
		\end{split}
	\end{equation}%
	A general approach is to consider a probability (pseudo)-metric $d(\cdot,\cdot)$ on the space of probability distributions on $\mathbb{R}^d$.
	Well-known examples of such a metric include the Kullback-Leibler divergence, the Kolmogorov-Smirnov distance, the Maximum Mean Discrepancy (MMD), etc.
	Observe that, under the null hypothesis, $d(\mathbb{P}_x, \mathbb{P}_z)=0$.
	The testing procedure consists in computing the empirical estimates $\widehat{\mathbb{P}}_x$ and $ \widehat{\mathbb{P}}_z$ and rejecting $H_0$ for ``large'' values of the statistics $d(\widehat{\mathbb{P}}_x, \widehat{\mathbb{P}}_z)$.
	This general formulation relies on a consistent estimation of arbitrary distributions from a finite number of samples.
	In the parametric setting, additional assumptions are made on the distribution functions: for instance, Gaussian assumption~\cite{Basseville1993,Chen2011a,Chen1997}, exponential family assumption~\cite{Frick2014,Adams2007}, etc.
	In the non-parametric setting, the distributions are only assumed to be continuous.
	They are not directly estimated; instead, the statistics $d(\widehat{\mathbb{P}}_x, \widehat{\mathbb{P}}_z)$ are computed~\cite{Liu2013,Clemencon2009,gretton2012kernel,Harchaoui2009}.\\
	In the context of \emph{single} change point detection, the two-sample test setting is adapted to assess whether a distribution change has occurred at some instant in the input signal.
	Practically, for a given index $t$, the homogeneity test is performed on the two populations $\{y_s\}_{s\leq t}$ and $\{y_s\}_{s>t}$.
	The estimated change point location is given by
	\begin{equation}
		\hat{t} = \operatorname{argmax}_t \ d(\widehat{\mathbb{P}}_{\bullet\leq t}, \widehat{\mathbb{P}}_{\bullet>t})
		\label{eq:binseg_discrepancy}
	\end{equation}
	where $\widehat{\mathbb{P}}_{\bullet\leq t}$ and $\widehat{\mathbb{P}}_{\bullet>t}$ are the empirical distributions of respectively $\{y_s\}_{s\leq t}$ and $\{y_s\}_{s>t}$.
\end{remark}%

%
%
\subsubsection{Binary segmentation}
\begin{figure}[t]
	\label{fig:BinSeg}
	\centering
	\includegraphics[width=\columnwidth]{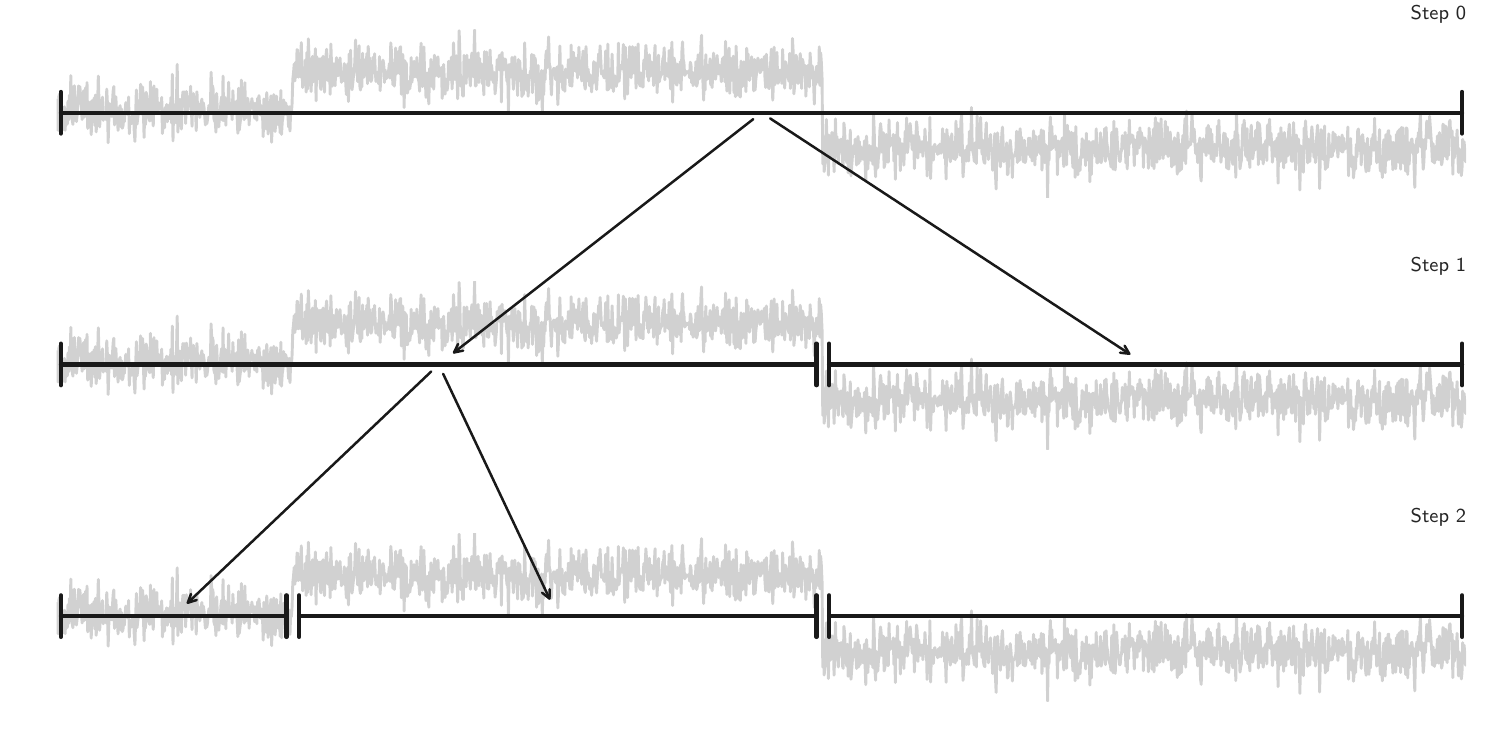}
	\caption{Schematic example of \texttt{BinSeg}}
	\label{fig:binseg_schema}
\end{figure}

Binary segmentation, denoted \texttt{BinSeg}, is a well-known alternative to optimal methods~\cite{Srivastava75binseg}, because it is conceptually simple and easy to implement~\cite{Olshen2004,Chen2011a, Killick2012a}.
\texttt{BinSeg} is a greedy sequential algorithm, outlined as follows.
The first change point estimate $\hat{t}^{(1)}$ is given by
\begin{equation}
	\hat{t}^{(1)} := \operatorname{argmin}_{1\leq t < T-1}\ \underbrace{c(y_{0..t}) + c(y_{t..T})}_{V(\mathcal{T}=\{t\})}.
	\label{eq:binseg_greedy}
\end{equation}
This operation is ``greedy'', in the sense that it searches the change point that lowers the most the sum of costs.
The signal is then split in two at the position of $\hat{t}^{(1)}$; the same operation is repeated on the resulting sub-signals until a stopping criterion is met.
A schematic view of the algorithm is displayed on Figure~\ref{fig:binseg_schema} and an implementation is given in Algorithm~\ref{alg:binseg}.
The complexity of \texttt{BinSeg} is of the order of $\mathcal{O}(T\log T)$.
This low complexity comes at the expense of optimality: in general, \texttt{BinSeg}'s output is only an approximation of the optimal solution.
As argued in~\cite{Bai1997,Killick2012a}, the issue is that the estimated change points $\hat{t}^{(k)}$ are not estimated from homogeneous segments and each estimate depends on the previous ones.
Change points that are close are imprecisely detected especially~\cite{Jandhyala2013}.
Applications of \texttt{BinSeg} range from financial time series~\cite{Chen1997,Chen2011a,Bai1997,Lavielle2007,fryzlewicz2014} to context recognition for mobile devices~\cite{989520} and array-based DNA copy number data~\cite{Picard2005, Olshen2004, Niu2012}.
\begin{algorithm}[h]
	\caption{Algorithm \texttt{BinSeg}}
	\begin{algorithmic}
		\State\textbf{Input:} signal $\{y_t\}_{t=1}^{T}$, cost function $c(\cdot)$, stopping criterion.
		\State Initialize $L\gets \{\;\} $. \Comment{Estimated breakpoints.}
		\Repeat
		\State $k\gets |L|$. \Comment{Number of breakpoints}
		\State $t_0\gets 0$ and $t_{k+1}\gets T$ \Comment{Dummy variables.}
		\If{$k>0$}
		\State Denote by $t_i$ ($i=1,\dots,k$) the elements (in ascending order) of $L$, ie $L=\{t_1,\dots,t_k\}$.
		\EndIf
		\State Initialize $G$ a $(k+1)$-long array. \Comment{list of gains}
		\For{$i=0,\dots,\,k$}
		\State $G[i] \gets c(y_{t_{i}..t_{i+1}}) - \min\limits_{t_{i}<t< t_{i+1}} [c(y_{t_{i}..t}) + c(y_{t..t_{i+1}})] $ .
		\EndFor
		\State $\widehat{i}\gets \operatorname{argmax}_i G[i]$
		\State $\hat{t} \gets \operatorname{argmin}\limits_{t_{\widehat{i}}<t< t_{\widehat{i}+1}}\ [c(y_{t_{\widehat{i}}..t}) + c(y_{t..t_{\widehat{i}+1}})]$.
		\State $L\gets L\cup\{\hat{t}\}$
		\Until{stopping criterion is met.}
		\State\textbf{Output:} set $L$ of estimated breakpoint indexes.
	\end{algorithmic}
	\label{alg:binseg}
\end{algorithm}
\paragraph{Related search methods.}
Several extensions of \texttt{BinSeg} have been proposed to improve detection accuracy.
\begin{itemize}
	\item[-]
	      Circular binary segmentation~\cite{Olshen2004} is a well-known extension of \texttt{BinSeg}.
	      This method is also a sequential detection algorithm that splits the original at each step.
	      Instead of searching for a single change point in each sub-signal, circular binary segmentation searches two change points.
	      Within each treated sub-segment, it assumes a so-called ``epidemic change model'': the parameter of interest shifts from one value to another at the first change point and returns to the original value at the second change point.
	      The algorithm is dubbed ``circular'' because, under this model, the sub-segment has its two ends (figuratively) joining to form a circle.
	      Practically, this method has been combined with $c_{L_2}$~\ref{eq:cost_mean_shift}, to detect changes in the mean of array-based DNA copy number data~\cite{Olshen2004,Lai2005,Willenbrock2005}.
	      A faster version of the original algorithm is described in~\cite{Venkatraman2007}.
	\item[-]
	      Another extension of \texttt{BinSeg} is the wild binary segmentation algorithm~\cite{fryzlewicz2014}.
	      In a nutshell, a single point detection is performed on multiple intervals with start and end points that are drawn uniformly.
	      Small segments are likely to contain at most one change but have lower statistical power, while the opposite is true for long segments.
	      After a proper weighting of the change score to account for the differences on sub-signals'	length, the algorithm returns the most ``pronounced'' ones, ie those that lower the most the sum of costs.
	      An important parameter of this method is the number of random sub-segments to draw.
	      Wild binary search is combined with $c_{L_2}$~\ref{eq:cost_mean_shift} to detect mean-shifts of univariate piecewise constant signals (up to 2000 samples)~\cite{fryzlewicz2014}.
\end{itemize}

\subsubsection{Bottom-up segmentation}
\begin{figure}[t]
	\label{fig:BotUp}
	\centering
	\includegraphics[width=\columnwidth]{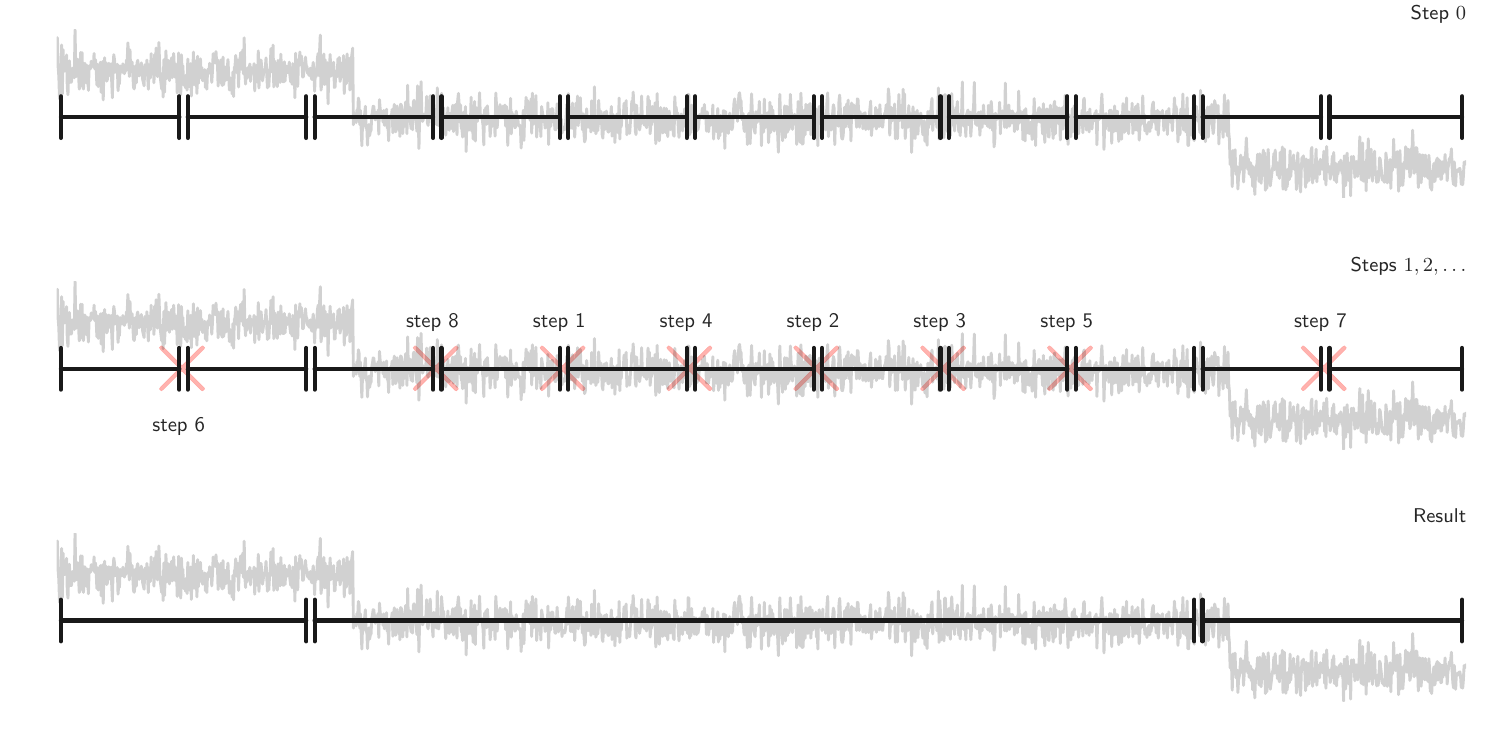}
	\caption{Schematic view of \texttt{BotUp}}
	\label{fig:botup_schema}
\end{figure}

Bottom-up segmentation, denoted \texttt{BotUp}, is the natural counterpart of \texttt{BinSeg}.
Contrary to \texttt{BinSeg}, \texttt{BotUp} starts by splitting the original signal in many small sub-signals and sequentially merges them until there remain only $K$ change points.
At every step, all potential change points (indexes separating adjacent sub-segments) are ranked by the discrepancy measure $d(\cdot,\cdot)$, defined in~\ref{eq:win_discrepancy}, between the segments they separate.
Change points with the lowest discrepancy are then deleted, meaning that the segments they separate are merged.
\texttt{BotUp} is often dubbed a ``generous'' method, by opposition to \texttt{BinSeg}, which is ``greedy''~\cite{Keogh2001}.
A schematic view of the algorithm is displayed on Figure~\ref{fig:botup_schema} and an implementation is provided in Algorithm~\ref{alg:botup}.
Its benefits are its linear computational complexity and conceptual simplicity.
However, if a true change point does not belong to the original set of indexes, \texttt{BotUp} never considers it.
Moreover, in the first iterations, the merging procedure can be unstable because it is performed on small segments, for which statistical significance is smaller.
In the literature, \texttt{BotUp} is somewhat less studied than its counterpart, \texttt{BinSeg}: no theoretical convergence study is available.
It has been applied on speech time series to detect mean and scale shifts~\cite{chen1998speaker}.
Besides, the authors of~\cite{Keogh2001} have found that \texttt{BotUp} outperforms \texttt{BinSeg} on ten different data sets such as physiological signals (ECG), financial time-series (exchange rate), industrial monitoring (water levels), etc.
\begin{algorithm}[t]
	\caption{Algorithm \texttt{BotUp}}
	\begin{algorithmic}
		\State\textbf{Input:} signal $\{y_t\}_{t=1}^{T}$, cost function $c(\cdot)$, stopping criterion, grid size $\delta>2$.
		\State Initialize $L\gets \{\delta, 2\delta,\dots, (\pa{T/\delta}-1)\,\delta\} $. \Comment{Estimated breakpoints.}
		\Repeat
		\State $k\gets |L|$. \Comment{Number of breakpoints}
		\State $t_0\gets 0$ and $t_{k+1}\gets T$ \Comment{Dummy variables.}
		\State Denote by $t_i$ ($i=1,\dots,k$) the elements (in ascending order) of $L$, ie $L=\{t_1,\dots,t_k\}$.
		\State Initialize $G$ a $(k-1)$-long array. \Comment{list of gains}
		\For{$i=1,\dots,\,k-1$}
		\State $G[i-1] \gets c(y_{t_{i-1}..t_{i+1}}) - [c(y_{t_{i-1}..t_{i}}) + c(y_{t_{i}..t_{i+1}})] $ .
		\EndFor
		\State $\widehat{i}\gets \operatorname{argmin}_i G[i]$
		\State Remove $t_{\widehat{i}+1}$ from $L$.
		\Until{stopping criterion is met.}
		\State\textbf{Output:} set $L$ of estimated breakpoint indexes.
	\end{algorithmic}
	\label{alg:botup}
\end{algorithm}

\section{Estimating the number of changes}\label{sec:review_unkn}
\begin{figure}[t]
	\centering
	\includegraphics[width=0.8\columnwidth]{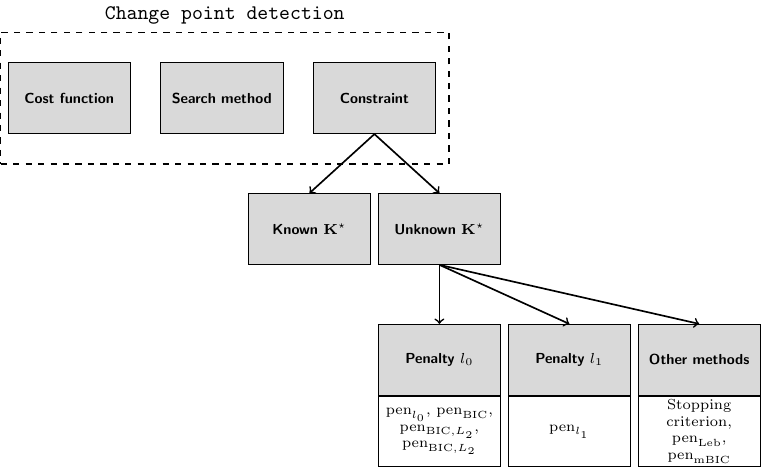}
	\caption{Typology of the constraints (on the number of change points) described in Section~\ref{sec:review_unkn}.}
	\label{fig:review-diagram-constraint}
\end{figure}

This section presents the third defining element of change detection methods, namely the constraint on the number of change points.
Here, the number of change points is assumed to be unknown \eqref{eq:review_unkn}.
Existing procedures are organized by the penalty function that they are based on.
Common heuristics are also described.
The organization of this section is schematically shown in Figure~\ref{fig:review-diagram-constraint}.
\subsection{Linear penalty}\label{sec:review-linear-pen}
Arguably the most popular choice of penalty~\cite{Killick2012a}, the linear penalty (also known as $l_0$ penalty) generalizes several well-known criteria from the literature such as the Bayesian Information Criterion (BIC) and the Akaike Information Criterion (AIC)~\cite{Yao1988, Yao1989}.
The linear penalty, denoted $\mbox{pen}_{l_0}$, is formally defined as follows.
\begin{pena}[$\mbox{pen}_{l_0}$]
	\label{def:review-pen-lin}
	The penalty function $\mbox{pen}_{l_0}$ is given by
	\begin{equation}
		\mbox{pen}_{l_0}(\mathcal{T}) := \beta |\mathcal{T}|
		\label{eq:unkn_l_0}
	\end{equation}
	where $\beta>0$ is the smoothing parameter.
\end{pena}
\noindent
Intuitively, the smoothing parameter controls the trade-off between complexity and goodness-of-fit (measured by the sum of costs): low values of $\beta$ favour segmentations with many regimes and high values of $\beta$ discard most change points.
\paragraph{Calibration.}
From a practical standpoint, once the cost function has been chosen, the only parameter to calibrate is the smoothing parameter.
Several approaches, based on model selection, can be found in the literature: they assume a model on the data, for instance~\eqref{eq:cost_pw_iid_model},~\eqref{eq:cost_linear_model},~\eqref{eq:cost_np_model}, and choose a value of $\beta$ that optimizes a certain statistical criterion.
The best-known example of such an approach is BIC, which aims at maximizing the constrained log-likelihood of the model.
The exact formulas of several linear penalties, derived from model selection procedures, are given the following paragraph.
Conversely, when no model is assumed, different heuristics are applied to tune the smoothing parameter.
For instance, one can use a procedure based on cross-validation~\cite{Arlot2010} or the slope heuristics~\cite{Birge2007}.
In~\cite{Rigailla,Truong2017}, supervised algorithms are proposed: the chosen $\beta$ is the one that minimizes an approximation of the segmentation error on an annotated set of signals.
\paragraph{Related penalties.}
A number of model selection criteria are special cases of the linear penalty $\mbox{pen}_{l_0}$.
For instance, under Model~\eqref{eq:cost_pw_iid_model} (iid with piecewise constant distribution), the constrained likelihood that is derived from the BIC and the penalized sum of costs are formally equivalent, upon setting $c=c_{\mbox{i.i.d.}}$ and $\mbox{pen} = \mbox{pen}_{\mbox{BIC}}$, where $\mbox{pen}_{\mbox{BIC}}$ is defined as follows.
\begin{pena}[$\mbox{pen}_{\mbox{BIC}}$]
	The penalty function $\mbox{pen}_{\mbox{BIC}}$ is given by
	\begin{equation}
		\mbox{pen}_{\mbox{BIC}} (\mathcal{T}) := \frac{p}{2} \log T \ |\mathcal{T}|
	\end{equation}
	where $p\geq1$ is the dimension of the parameter space in~\eqref{eq:cost_pw_iid_model}.
\end{pena}
\noindent
In the extensively studied model of an univariate Gaussian signal, with fixed variance $\sigma^2$ and piecewise constant mean, the penalty $\mbox{pen}_{\mbox{BIC}}$ becomes $\mbox{pen}_{L_2}$, defined below.
Historically, it was one of the first penalties introduced for change point detection~\cite{Schwarz1978,Yao1988}.
\begin{pena}[$\mbox{pen}_{\mbox{BIC},L_2}$]
	The penalty function $\mbox{pen}_{\mbox{BIC},L_2}$ is given by
	\begin{equation}
		\mbox{pen}_{\mbox{BIC},L_2} (\mathcal{T}) :=\sigma^2\log T\ |\mathcal{T}|.
	\end{equation}
	where $\sigma$ is the standard deviation and $T$ is the number of samples.
\end{pena}
\noindent
In the same setting, AIC, which is a generalization of Mallows' $C_p$~\cite{Mallows1973}, also yields a linear penalty, namely $\mbox{pen}_{\mbox{AIC},L_2}$, defined as follows.
\begin{pena}[$\mbox{pen}_{\mbox{AIC},L_2}$]
	The penalty function $\mbox{pen}_{\mbox{AIC},L_2}$ is given by
	\begin{equation}
		\mbox{pen}_{\mbox{AIC},L_2} (\mathcal{T}) :=\sigma^2\ |\mathcal{T}|.
	\end{equation}
	where $\sigma$ is the standard deviation.
\end{pena}
%
\subsection{Fused lasso}
For the special case where the cost function is $c_{L_2}$, a faster alternative to $\mbox{pen}_{l_0}$ can be used.
To that end, the $l_0$ penalty is relaxed to a $l_1$ penalty~\cite{Harchaoui2010,Vert2010}.
The resulting penalty function, denoted $\mbox{pen}_{l_1}$, is defined as follows.
\begin{pena}[$\mbox{pen}_{l_1}$]
	The penalty function $\mbox{pen}_{l_1}$ is given by
	\begin{equation}
		\mbox{pen}_{l_1} (\mathcal{T}) := \beta \sum_{k=1}^{|\mathcal{T}|} \norm{\bar{y}_{t_{k-1}..t_{k}} - \bar{y}_{t_{k}..t_{k+1}}}_1
	\end{equation}
	where $\beta>0$ is the smoothing parameter, the $t_k$ are the elements of $\mathcal{T}$ and $\bar{y}_{t_{k-1}..t_{k}}$ is the empirical mean of sub-signal $y_{t_{k-1}..t_{k}}$.
\end{pena}
\noindent
This relaxation strategy (from $l_0$ to $l_1$) is shared with many developments in machine learning, for instance sparse regression, compressive sensing, sparse PCA, dictionary learning~\cite{Hastie2009}, where $\mbox{pen}_{l_1}$ is also referred to as the fused lasso penalty.
In numerical analysis and image denoising, it is also known as the total variation regularizer~\cite{Harchaoui2010,Seichepine2014,Vert2010}.
Thanks to this relaxation, the optimization of the penalized sum of costs~\eqref{eq:review_sum_of_cost} in \eqref{eq:review_unkn} is transformed into a convex optimization problem, which can be solved efficiently using Lars (for ``least absolute shrinkage and selection operator'')~\cite{Harchaoui2010,Vert2010}.
The resulting complexity is of this order of $\mathcal{O}(T\log T)$~\cite{Tibshirani1996,Hastie2009}.
From a theoretical standpoint, under the mean-shift model (piecewise constant signal with Gaussian white noise), the estimated change point fractions are asymptotically consistent~\cite{Harchaoui2010}.
This result is demonstrated for an appropriately converging sequence of values of $\beta$.
This consistency property is obtained even though classical assumptions from the Lasso regression framework (such as the irrepresentable condition) are not satisfied~\cite{Harchaoui2010}.
In the literature, $\mbox{pen}_{l_1}$, combined with $c_{L_2}$, is applied on DNA sequences~\cite{Vert2010, Hocking2013}, speech signals~\cite{Angelosante2012} and climatological data~\cite{Jeon2016}.
\subsection{Complex penalties}
Several other penalty functions can be found in the literature.
However they are more complex, in the sense that the optimization of the penalized sum of cost is not tractable.
In practice, the solution is found by computing the optimal segmentations with $K$ change points, with $K=1,2,\dots,K_{\max}$ for a sufficiently large $K_{\max}$, and returning the one that minimizes the penalized sum of costs.
When possible, the penalty can also be approximated by a linear penalty, in which case, \texttt{Pelt} can be used.
In this section, we describe two examples of complex penalties.
Both originate from theoretical considerations, under the univariate mean-shift model, with the cost function $c_{L_2}$.
The first example is the modified BIC criterion (mBIC)~\cite{Zhang2007}, which consists in maximizing the asymptotic posterior probability of the data.
The resulting penalty function, denoted $\mbox{pen}_{\mbox{mBIC}}$, depends on the number and repartition of the change point indexes: intuitively, it favours evenly spaced change points.
\begin{pena}[$\mbox{pen}_{\mbox{mBIC}}$]
	The penalty function $\mbox{pen}_{\mbox{mBIC}}$ is given by
	\begin{equation}
		\mbox{pen}_{\mbox{mBIC}} (\mathcal{T}) := 3|\mathcal{T}|\log T + \sum_{k=0}^{|\mathcal{T}|+1} \log(\frac{t_{k+1}-t_{k}}{T})
	\end{equation}
	where the $t_k$ are the elements of $\mathcal{T}$.
\end{pena}
\noindent
In~\cite{Lebarbier2005}, a model selection procedure leads to another complex penalty function, namely $\mbox{pen}_{\mbox{Leb}}$.
Upon using this penalty function, the penalized sum of costs satisfied a so-called oracle inequality, which holds in a non-asymptotic setting, contrary to the other penalties previously described.
\begin{pena}[$\mbox{pen}_{\mbox{Leb}}$]
	The cost function $\mbox{pen}_{\mbox{Leb}}$ is given by
	\begin{equation}
		\mbox{pen}_{\mbox{Leb}} (\mathcal{T}) := \frac{|\mathcal{T}|+1}{T}\sigma^2(a_1 \log \frac{|\mathcal{T}|+1}{T} + a_2)
	\end{equation}
	where $a_1>0$ and $a_2>0$ are positive parameters and $\sigma^2$ is the noise variance.
\end{pena}
\noindent

\section{Summary table}\label{sec:review_summary}
This literature review is summarized in Table~\ref{tab:summary}.
When applicable, each publication is associated with a search method (such as \texttt{Opt}, \texttt{Pelt}, \texttt{BinSeg} or \texttt{Win}); this is a rough categorization rather than an exact implementation.
Note that \texttt{Pelt} (introduced in~\citeyear{Killick2012a}) is sometimes associated with publications prior to 2012.
It is because some linear penalties~\cite{Mallows1973,Zhang2007} were introduced long before \texttt{Pelt} was, and authors then resorted to quadratic (at best) algorithms.
Nowadays, the same results can be obtained faster with \texttt{Pelt}.
A guide of computational complexity is also provided.
Quadratic methods are the slowest and have only one star while linear methods are given three stars.
Algorithms for which the number of change points is an explicit input parameter work under the ``known $K$'' assumption.
Algorithms that can be used even if the number of change points is unknown work under the ``unknown $K$'' assumption.
(Certain methods can accommodate both situations.)

\begin{landscape}
	\thispagestyle{empty}%
	\begin{table}[p]
		\centering\tiny
		\setlength\extrarowheight{0.5em}
		\begin{tabular}{cccccccc}
			\hline
			{Publication}                                        & {Search method} & {Cost function}             & \multicolumn{2}{c}{Known $K$} & {Scalability (wrt $T$)} & {Package}          & {Additional information}                                                                                         \\
			                                                     &                 &                             & Yes                           & No                       &                    &                          &                                                                                       \\ \hline
			\mycite{Srivastava75binseg}, \mycite{Vostrikova1981} & \texttt{BinSeg}       & $c_{L_2}$                   & \ding{51}                        & -                        & \ding{72}\ding{72}\ding{72} & \ding{51}                   &                                                                                       \\
			\mycite{Yao1988}                                     & \texttt{Opt}          & $c_{L_2}$                   & -                             & \ding{51}                   & \ding{72}\ding{73}\ding{73} & -                        & Bayesian information criterion (BIC)                                                  \\
			\mycite{Basseville1993}                              & \texttt{Opt}          & $c_{\mbox{i.i.d.}},c_{L_2}$ & -                             & -                        & \ding{72}\ding{72}\ding{72} & -                        & single change point                                                                   \\
			\mycite{Bai1994}, \mycite{bai2003computation}        & \texttt{Opt}          & $c_{\mbox{linear},L_2}$     & -                             & -                        & \ding{72}\ding{72}\ding{73} & -                        & single change point                                                                   \\
			\mycite{Bai1995}                                     & \texttt{Opt}          & $c_{\mbox{linear},L_1}$     & -                             & -                        & \ding{72}\ding{72}\ding{73} & -                        & single change point                                                                   \\
			\mycite{Lavielle1998}                                & \texttt{Opt}          & $c_{AR}$                    & \ding{51}                        & -                        & \ding{72}\ding{73}\ding{73} & -                        &                                                                                       \\
			\mycite{Bai2000}                                     & \texttt{Opt}          & $c_{AR}$                    & \ding{51}                        & -                        & \ding{72}\ding{73}\ding{73} & -                        &                                                                                       \\
			\mycite{Birge2001}, \mycite{Birge2007}               & \texttt{Opt}          & $c_{L_2}$                   & -                             & \ding{51}                   & \ding{72}\ding{73}\ding{73} & -                        & model selection                                                                       \\
			\mycite{Bai2003}                                     & \texttt{Opt}          & $c_{L_2}$                   & \ding{51}                        & -                        & \ding{72}\ding{73}\ding{73} & -                        &                                                                                       \\
			\mycite{Olshen2004}, \mycite{Venkatraman2007}        & \texttt{BinSeg}       & $c_{L_2}$                   & \ding{51}                        & \ding{51}                   & \ding{72}\ding{72}\ding{72} & \ding{51}                   &                                                                                       \\
			\mycite{Lebarbier2005}                               & \texttt{Opt}          & $c_{L_2}$                   & -                             & \ding{51}                   & \ding{72}\ding{73}\ding{73} & -                        & model selection                                                                       \\
			\mycite{desobry2005online}                           & \texttt{Win}          & $c_{kernel}$                & -                             & \ding{51}                   & \ding{72}\ding{72}\ding{72} & -                        & dissimilarity measure (one-class SVM), see Remark~\ref{rem:win_two_sample_test}       \\
			\mycite{harchaoui2007retrospective}                  & \texttt{Opt}          & $c_{kernel},c_{rbf}$        & \ding{51}                        & -                        & \ding{72}\ding{73}\ding{73} & -                        &                                                                                       \\
			\mycite{Zhang2007}                                   & \texttt{Pelt}         & $c_{L_2}$                   & -                             & \ding{51}                   & \ding{72}\ding{72}\ding{73} & -                        & modified BIC                                                                          \\
			\mycite{Harchaoui2009}                               & \texttt{Win}          & -                           & \ding{51}                        & \ding{51}                   & \ding{72}\ding{72}\ding{72} & -                        & dissimilarity measure (Fisher discriminant), see Remark~\ref{rem:win_two_sample_test} \\
			\mycite{Levy-Leduc2009}, \mycite{Lung-Yut-Fong2012}  & \texttt{Win}          & $c_{\mbox{rank}}$           & \ding{51}                        & \ding{51}                   & \ding{72}\ding{72}\ding{72} & \ding{51}                   & dissimilarity measure (rank-based), see Remark~\ref{rem:win_two_sample_test}          \\
			\mycite{Bai2010}                                     & \texttt{Opt}          & $c_{L_2},c_{\Sigma}$        & -                             & -                        & \ding{72}\ding{72}\ding{73} & -                        & single change point                                                                   \\
			\mycite{Vert2010}                                    & Fused Lasso     & $c_{L_2}$                   & -                             & \ding{51}                   & \ding{72}\ding{72}\ding{72} & -                        & Tikhonov regularization                                                               \\
			\mycite{Harchaoui2010}                               & Fused Lasso     & $c_{L_2}$                   & -                             & \ding{51}                   & \ding{72}\ding{72}\ding{72} & -                        & total variation regression ($\mbox{pen}_{l_1}$)                                       \\
			\mycite{arlot2012kernel}                             & \texttt{Opt}          & $c_{kernel},c_{rbf}$        & \ding{51}                        & \ding{51}                   & \ding{72}\ding{73}\ding{73} & -                        &                                                                                       \\
			\mycite{Killick2012a}                                & \texttt{Pelt}         & any $c(\cdot)$              & -                             & \ding{51}                   & \ding{72}\ding{72}\ding{73} & \ding{51}                   &                                                                                       \\
			\mycite{Angelosante2012}                             & Fused Lasso     & $c_{AR}$                    & -                             & \ding{51}                   & \ding{72}\ding{72}\ding{72} & -                        & Tikhonov regularization                                                               \\
			\mycite{Liu2013}                                     & \texttt{Win}          & -                           & -                             & \ding{51}                   & \ding{72}\ding{72}\ding{72} & -                        & dissimilarity measure (density ratio), see Remark~\ref{rem:win_two_sample_test}       \\
			\mycite{Rigailla}                                    & \texttt{Pelt}         & $c_{L_2}$                   & -                             & \ding{51}                   & \ding{72}\ding{72}\ding{73} & -                        & supervised method to learn a penalty level ($\mbox{pen}_{l_0}$)                       \\
			\mycite{fryzlewicz2014}                              & \texttt{BinSeg}       & $c_{L_2}$                   & \ding{51}                        & \ding{51}                   & \ding{72}\ding{72}\ding{72} & \ding{51}                   & univariate signal                                                                     \\
			\mycite{Lajugie2014}                                 & \texttt{Opt}          & $c_{M}$                     & \ding{51}                        & -                        & \ding{72}\ding{73}\ding{73} & -                        & supervised method to learn a suitable metric                                          \\
			\mycite{Frick2014}                                   & \texttt{BinSeg}       & $c_{\mbox{i.i.d.}}$         & \ding{51}                        & \ding{51}                   & \ding{72}\ding{72}\ding{72} & \ding{51}                   & exponential distributions family                                                      \\
			\mycite{Lung-Yut-Fong2015}                           & \texttt{Opt}          & $c_{\mbox{rank}}$           & \ding{51}                        & -                        & \ding{72}\ding{73}\ding{73} & \ding{51}                   &                                                                                       \\
			\mycite{Garreau2016}                                 & \texttt{Pelt}         & $c_{kernel},c_{rbf}$        & \ding{51}                        & \ding{51}                   & \ding{72}\ding{73}\ding{73} & -                        &                                                                                       \\
			\mycite{Haynes2017}                                  & \texttt{Pelt}         & any $c(\cdot)$              & -                             & \ding{51}                   & \ding{72}\ding{72}\ding{73} & -                        &                                                                                       \\
			\mycite{Chakar2017}                                  & \texttt{Pelt}         & $c_{AR}$                    & \ding{51}                        & \ding{51}                   & \ding{72}\ding{73}\ding{73} & \ding{51}                   &
		\end{tabular}
		\caption{Summary table of literature review.}
		\label{tab:summary}
	\end{table}
\end{landscape}

\section{Presentation of the Python package}
\label{sec:python}

Most of the approaches presented in this article are included in a Python scientific library for multiple change point detection in multivariate signals called \texttt{ruptures} \cite{packrup}. The \texttt{ruptures} library is written in pure Python and available on Mac OS X, Linux and Windows platforms. Source code is available from \cite{packrup} under the BSD license and deployed with  a complete documentation that includes installation instructions and explanations with code snippets on advance use.

\begin{figure}[t]
	\centering
	\includegraphics[width=0.8\columnwidth]{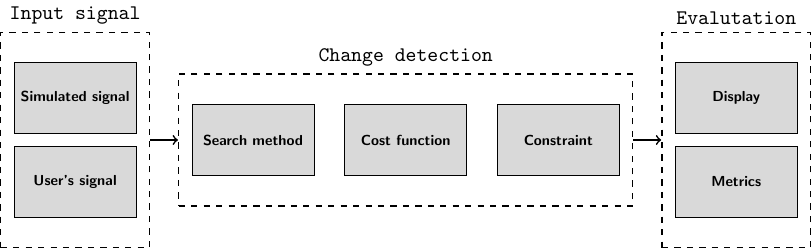}
	\caption{Schematic view of the \texttt{ruptures} package.}
	\label{fig:diagram}
\end{figure}

A schematic view is displayed on Figure~\ref{fig:diagram}.
Each block of this diagram is described in the following brief overview of \texttt{ruptures}' features.

\begin{itemize}
	\item\textbf{Search methods}\enspace
	Our package includes the main algorithms from the literature, namely dynamic programming, detection with a $l_{0}$ constraint, binary segmentation, bottom-up segmentation and window-based segmentation. 
	This choice is the result of a trade-off between exhaustiveness and adaptiveness.
	Rather than providing as many methods as possible, only algorithms which have been used in several different settings are included.
	In particular, numerous ``mean-shift only'' detection procedures were not considered.
	Implemented algorithms have sensible default parameters that can be changed easily through the functions' interface.
	\item\textbf{Cost functions}\enspace 
	Cost functions are related to the type of change to detect. 
	Within \texttt{ruptures}, one has access to parametric cost functions that can detect shifts in standard statistical quantities (mean, scale, linear relationship between dimensions, autoregressive coefficients, etc.) and non-parametric cost functions (kernel-based or Mahalanobis-type metric) that can, for instance, detect distribution changes~\citep{harchaoui2007retrospective,Lajugie2014}.
	\item\textbf{Constraints}\enspace
	All methods can be used whether the number of change points is known or not.
	In particular, \texttt{ruptures} implements change point detection under a cost budget and with a linear penalty term~\citep{Killick2012a, Maidstone2017}.
	\item\textbf{Evaluation}\enspace 
	Evaluation metrics are available to quantitatively compare segmentations, as well as a display module to visually inspect algorithms' performances.
	\item\textbf{Input}\enspace 
	Change point detection can be performed on any univariate or multivariate signal that fits into a \emph{Numpy} array.
	A few standard non-stationary signal generators are included.
	\item\textbf{Consistent interface and modularity}\enspace Discrete optimization methods and cost functions are the two main ingredients of change point detection.
	Practically, each is related to a specific object in the code, making the code highly modular: available optimization methods and cost functions can be connected and composed.
	An appreciable by-product of this approach is that a new contribution, provided its interface follows a few guidelines, can be integrated seamlessly into \texttt{ruptures}.
	\item\textbf{Scalability}\enspace
	Data exploration often requires to run several times the same methods with different sets of parameters.
	To that end, a cache is implemented to keep intermediate results in memory, so that the computational cost of running the same algorithm several times on the same signal is greatly reduced.
	We also add the possibility for a user with speed constraints to sub-sample their signals and set a minimum distance between change points. 
\end{itemize}

\section{Conclusion}
In this article, we have reviewed numerous methods to perform change point detection, organized within a common framework.
Precisely, all methods are described as a collection of three elements: a cost function, a search method and a constraint on the number of changes to detect.
This approach is intended to facilitate prototyping of change point detection methods: for a given segmentation task, one can pick among the described elements to design an algorithm that fits its use-case.
Most detection procedures described above are available within the Python language from the package \texttt{ruptures} \cite{packrup}, which is the most comprehensive change point detection library.
Its consistent interface and modularity allow painless comparison between methods and easy integration of new contributions.
In addition, a thorough documentation is available for novice users.
Thanks to the rich Python ecosystem, \texttt{ruptures} can be used in coordination with numerous other scientific libraries .

%
\clearpage

\singlespacing
\footnotesize
\bibliographystyle{unsrtnat}
\bibliography{biblio}
%
\end{document}